\documentclass[%
 reprint,
superscriptaddress,
 amsmath,amssymb,
 aps,
]{revtex4-2}
\usepackage{graphicx}
\usepackage{dcolumn}
\usepackage{bm}

\usepackage{lipsum}
\usepackage{xcolor}
\usepackage[export]{adjustbox}
\begin{document}

\preprint{APS/123-QED}

\title{All optical chaos synchronization between nonidentical optomechanical cavities}
\author{Souvik Mondal}
 \affiliation{%
Department of Electronics and Electrical Communication Engineering, IIT Kharagpur, West Bengal, 721302, India}%
\author{Murilo S. Baptista}
 \affiliation{%
School of Natural and Computing Sciences, University of Aberdeen, Aberdeen AB24 3UE, United Kingdom}%
 \author{Kapil Debnath}%
 \email{kapil.debnath@abdn.ac.uk}
 \affiliation{%
School of Natural and Computing Sciences, University of Aberdeen, Aberdeen AB24 3UE, United Kingdom}%
\begin{abstract}
Optomechanical cavities, with nonlinear photon-phonon interactions, offer a more compact approach to chaos generation than conventional feedback-based optical systems. However, proper study on chaos synchronization of two optomechanical cavities connected by optical means  is still unexplored. In this work, we theoretically investigate all-optical complete synchronization between unidirectionally coupled optomechanical cavities in the master-slave configuration. Traditionally, achieving complete synchronization in nonlinear coupled oscillators and in optical systems necessitates identical systems.
Our findings, which arise naturally from the fundamental mathematical properties of optomechanical cavities, demonstrate that parameter heterogeneity can, in fact, not only enable complete synchronization but make it stable.
\end{abstract}
\maketitle
\section{Introduction}\label{sec1}
The study of nonlinear dynamics and chaos has significantly advanced science and technology, offering fresh insights across various fields such as biology, chemistry, physics and many others \cite{strogatz2018nonlinear,marco2010}. These advances have facilitated better understanding of complex dynamics that exhibits unpredictable behaviours \cite{ott1981strange}. From a technological standpoint, a pivotal moment in this scientific journey occurred in 1990, when Pecora and Carroll demonstrated that even chaotic systems possessing sensibility to the initial conditions and generating unpredictable beahviour could achieve complete synchronization, a concept traditionally associated with order \cite{PhysRevLett.64.821}. The discovery of chaos synchronization has since become a cornerstone for innovative applications in secure communications \cite{jovic2011chaotic}, cryptography \cite{mishkovski2011chaos}, distributed sensor networks \cite{zhang2024chaos} and artificial neural networks \cite{aihara1990chaotic}.

Among various platforms, optical systems, with their inherent advantages, such as wide bandwidth, higher dimensionality, and seamless integration with existing fiber optic networks are increasingly recognized as the ideal platform for demonstrating chaos synchronization and its applications \cite{soriano2013complex,ohtsubo2017semiconductor}. The optical system consists of a semiconductor laser with delayed feedback where the feedback is realized by purely optical or electro-optical in nature. Optomechanical cavities (OMCs), by using the inherent nonlinear interaction between light fields and mechanical oscillations, provide a fundamentally different approach to chaos generation which does not require any feedback loops \cite{carmon2007chaotic,carmon2005temporal,metzger2008self,marino2011chaotically,hollander2012self,marino2013coexisting, wu2017mesoscopic,navarro2017nonlinear}. Therefore, the system has the potential to provide more scalability and better integrability in chips than the semiconductor lasers; as well as controlling of the optomechanical nonlinearities \cite{gil2017light,sciamanna2016vibrations,ren2022topological}. While progress has been made in the study and generation of chaos in OMCs \cite{mondal2024chaotic,zhang2020intermittent,lu2015pt,ma2014formation}, there has been limited study based on chaos synchronization. The study in \cite{yang2019chaotic, monifi2016optomechanically} demonstrates chaos synchronization between two weakly driven optical modes mediated by a common mechanical mode and this same mechanical mode is coupled to another optical mode with strong optical pump to induce chaos. Another study in \cite{madiot2021bichromatic} demonstrates that bichromatic electrical actuation drives two mechanical normal modes in the chaotic regime and brings about synchronization between those modes in which their amplitudes and phases are accessed via optical readout. These results suggest approaches to synchronization mediated by mechanical motions. However, we are intrigued by the possibility of achieving synchronization by optical means and thereby provides motivation to theoretically examine all-optical chaos synchronization between two separate OMCs connected by an optical fiber or optical waveguide.
\begin{figure*}[t]
    \centering
    \includegraphics[width=0.85\linewidth]{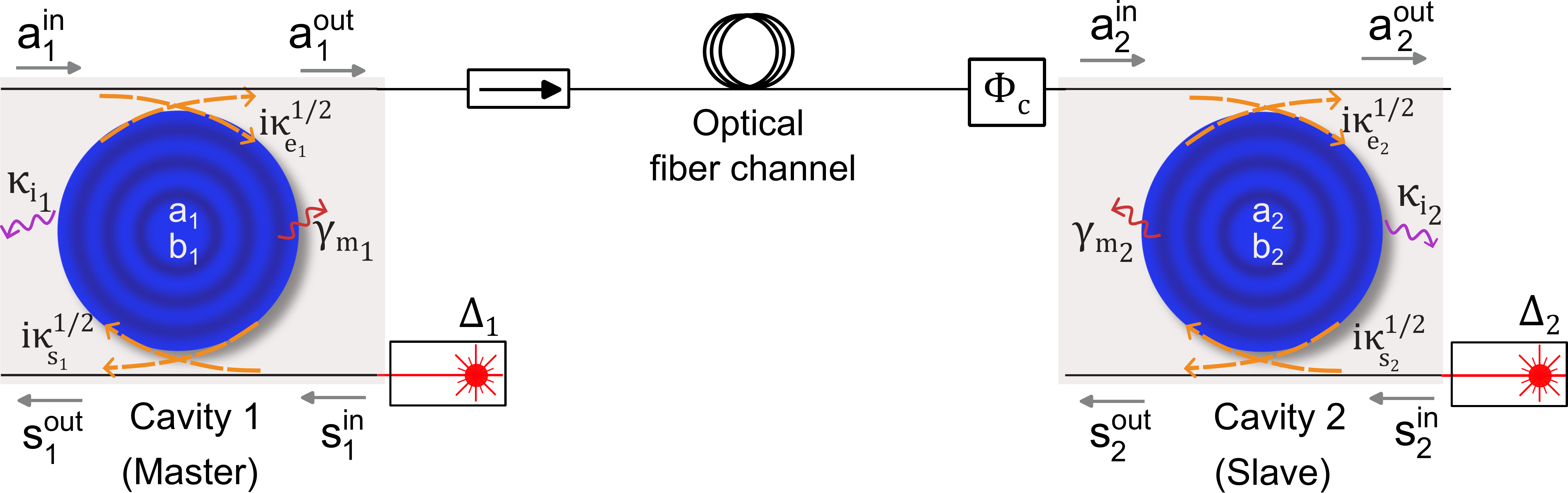}
    \caption{Schematic of a master and a slave OMC coupled unidirectionally by optical fiber with the excitation laser being provided separately with amplitude $s^{\text{in}}_{1,2}$ and detuning $\Delta_{1,2}$. The phase controller block ``$\Phi_c$" adjusts the phase of the optical field coming from master cavity to realize chaos synchronization}
    \label{fig1}
\end{figure*}

The synchronization of chaotic dynamics occurs in several forms \cite{pikovsky2001synchronization,boccaletti2002synchronization,eroglu2017} and, among them, our particular interest lies in studying complete chaos synchronization in OMCs, as it is the ``strongest" form of synchronization where the state of the systems are exactly the same \cite{boccaletti2002synchronization}. In the conventional optical system as well as in general coupled nonlinear oscillators, such a form of synchronization requires the system parameters to be identical or nearly identical \cite{locquet2002synchronization,murakami2002synchronization,eichler2018semiconductor,acharyya2012synchronization}, which poses a challenging task in practical implementation. So it is desirable for a system to achieve strong synchronization which are robust against variation of parameters. Generally, chaos synchronization in nonidentical oscillators  induce weaker forms of synchronization, such as generalized synchronization \cite{rulkov1995generalized}, cluster synchronization \cite{dahms2012cluster,belykh2003persistent,pecora2014cluster} or phase synchronization \cite{zhou2002noise,kiss2002phase}.In fact similar weak synchronization persists for optomechanical system mentioned in \cite{yang2019chaotic,madiot2021bichromatic} for mismatched parameter values. Complete synchronization can be made to achieve between different chaotic oscillators through methods such as active control technique \cite{chen2005synchronization,yassen2005chaos}, but the chaotic oscillators are primarily electronic based system.  In laser based optical system, the phenomenon of complete synchronization in non-identical elements is present in large networks, involving three or more lasers \cite{williams2013experimental,landsman2007complete,zamora2010crowd}. Therefore, we also want to explore the possibility of achieving stable complete synchronization that can exist in a two coupled oscillator configuration of all-optical master-slave OMC with parameter heterogeneity.  An additional phase controller element  is needed in this configuration to achieve our desired results. Furthermore, the insights of chaos synchronization is explained through quantities calculated by the positive Lyapunov exponents and that provide estimates for upper bounds for the Kolmogorov-Sinai entropy and mutual information rate, and provide the pathway to complete synchronization as the coupling between the OMCs increases, determining regimes for which the synchronization manifold is stable. 

The rest of the paper is organized as follows. In Sec \ref{sec2} we describe the dynamical equations of the coupled system along with the discussion on the conditions, stability and various metrics of synchronization. The numerical results are shown in Sec. \ref{sec3} for single OMC characterizing various nonlinear regimes. The phenomenon of chaos synchronization is described in detail in Sec. \ref{sec4} . In Sec. \ref{sec5} we discussed the behaviour of the quality of chaos synchronization against different parameters. We then discussed briefly about possible experimental realization in Sec. \ref{sec6}. Finally we summarize our work in Sec. \ref{sec7}.

\section{System Modeling}\label{sec2}
This section presents the mathematical model of the system, derives the relations among system parameters and identifies different metrics to understand complete chaos synchronization. 

\subsection{Mathematical description of the coupled OMC}
Our system,  as shown in Fig. \ref{fig1}, consists of two micro-toroid based OMCs \cite{kippenberg2005analysis,schliesser2008high,weis2010optomechanically} which are unidirectionally coupled from cavity 1 (master) to cavity 2 (slave) by an optical fiber of efficiency $\eta$.  We consider that each of the cavities has a dimensionless and complex optical mode amplitude $a_{1,2}$ and  mechanical mode amplitude $b_{1,2}$, where the subindex ``1" represents master, and ``2" represents the slave. The intrinsic loss rate of the optical mode in the cavities is given by $\kappa_{i_{1,2}}$ while $\gamma_{m_{1,2}}$ is the loss rate corresponding to mechanical mode. The external coupling rate between the cavities and the optical fiber (or waveguide) channel is given by $\kappa_{e_{1,2}}$. We consider the cavities to be excited by continuous wave lasers with constant amplitude $s^{\text{in}}_{1,2}$ and detuning $\Delta_{1,2}$ through separate waveguide-cavity coupling with coupling rate $\kappa_{s_{1,2}}$. The detuning is given by the difference between laser frequency ($\omega_L$) and the optical resonance frequency ($\omega_{c_{1,2}}$), which is, $\Delta_{1,2}=\omega_L-\omega_{c_{1,2}}$. The input (output) optical field of the cavities through the optical fiber channel is labeled by $a^{\text{in}}_{1,2}$ ($a^{\text{out}}_{1,2}$). These fields are related by $a_{1(2)}^{\text{out}}(t)=i\sqrt{\kappa_{e_{1(2)}}}a_{1(2)}(t)+a_{1(2)}^{\text{in}}(t)$ through input-output formalism of optical cavities \cite{walls2008input}. Similarly the output fields of the cavities at their respective driving waveguide are given as $s_{1(2)}^{\text{out}}(t)=i\sqrt{\kappa_{s_{1(2)}}}a_{1(2)}(t)+s_{1(2)}^{\text{in}}$. The block labeled by ``$\Phi_c$" is the phase controller block which adjusts the phase of the propagation delayed optical field $a^{\text{out}}_1$ and also locks to the desired value; and thereby the resultant field is fed into cavity 2. This block serves an important role in synchronization as discussed later and this can be realized by using optical phase-locking scheme \cite{ristic2009optical,arafin2017heterodyne}.
Assuming the propagating distance to be small, the effect of attenuation in the fiber should also be minimum and thereby the efficiency $\eta$ can be set to $1$. The semiclassical equation of motions of the variables $a_{1,2}$ and $b_{1,2}$, in the rotating frame of driving laser, are written as \cite{aspelmeyer2014cavity}
 \begin{subequations}
 \begin{flalign}
     \frac{\text d a_{1,2}}{\text d t}=&\left(i\Delta_{1,2}-\frac{\kappa_{1,2}}{2}\right)a_{1,2}-i2g_{0_{1,2}}\text{Re}\left(b_{1,2}\right)a_{1,2}\\\nonumber&+i\sqrt{\kappa_{s_{1,2}}}s^{\text{in}}_{1,2}+i\sqrt{\kappa_{e_{1,2}}}a^{\text{in}}_{1,2}+\sqrt{\kappa_{1,2}}\xi_{a_{1,2}}(t),
     \\
     \frac{\text d b_{1,2}}{\text d t}=&-\left(i\omega_{m_{1,2}}+\frac{\gamma_{m_{1,2}}}{2}\right)b_{1,2}-ig_{0_{1,2}}|a_{1,2}|^2\\\nonumber&+\sqrt{\gamma_{m_{1,2}}}\xi_{b_{1,2}}(t),
 \end{flalign}
 \label{eq1}
\end{subequations}

\noindent where $g_{0_{1,2}}$ is the optomechanical coupling rate and $\omega_{m_{1,2}}$ is the resonance frequency of the mechanical modes in both the cavities. The total loss rate of the optical mode in each of the cavity is given by $\kappa_{1(2)}=\kappa_{i_{1(2)}}+\kappa_{s_{1(2)}}+\kappa_{e_{1(2)}}$. The phase $+i$ of the coupling term $\sqrt{\kappa_{e_{1,2}}}$ and $\sqrt{\kappa_{s_{1,2}}}$ in the dynamical equation is arbitrarily chosen based on the choice of reference plane \cite{784592}. $\xi_{a_{1,2}}$ and $\xi_{b_{1,2}}$ are the white Gaussian complex noise processes of optical and mechanical modes respectively, with mean value of zero and satisfy the correlation property,
\begin{subequations}
\begin{eqnarray}
    \langle\xi_{a_i}^{\ast}(t)\xi_{a_j}(t')\rangle=\left(\bar n_{a_i}+\frac{1}{2}\right)\delta_{ij}\delta(t-t')\;
\\
\langle\xi_{b_i}^{\ast}(t)\xi_{b_j}(t')\rangle=\left(\bar n_{b_i}+\frac{1}{2}\right)\delta_{ij}\delta(t-t'), 
\end{eqnarray}
\label{eq2}
\end{subequations}
in which $i,j=\{1,2\}$ are the subscripts denoting cavity 1 and cavity 2. $\bar n_{a_{1,2}} (\bar n_{b_{1,2}})$ are the mean thermal excitation number of the optical (mechanical) modes of the cavities given by $\bar n_{a_{1,2}}(\bar n_{b_{1,2}})\approx k_BT/\hbar\omega_{c_1,2} \left(k_BT/\hbar\omega_{m_1,2}\right)$ at temperature $T$ (in Kelvin). The effect of optical mode noise $\xi_{a_{1,2}}$ can  be safely omitted in our numerical study in the subsequent sections. This is because of the high resonance frequency of the cavity in the order of $10^{14}$ Hz which makes the thermal excitation number $\bar n_{a_{1,2}}\ll1$ at finite temperature $T$. For further simplification of study, we assume the system is operated in the low temperature, which can be achieved through the use of dilute refrigerator, such that the mean thermal excitation number of the mechanical mode $\bar n_{b_{1,2}}$ can be neglected as well \cite{mckenna2019cryogenic,kuhn2014free} and therefore $\xi_{b_{1,2}}$ is omitted.  Also the effect of vacuum field fluctuations on the optical and mechanical mode amplitude can be ignored as operable energy in the cavity is high in chaotic regime. In addition, the driving laser sources are assumed to exhibit low noise, which can be achieved by using lasers with narrow linewidth in the range of few kilohertz or less \cite{fu2017review,yang2020chaotic}. Now, for convenience, we consider that the mechanical resonance frequencies and optomechanical coupling rates are similar, that is, $\omega_{m_1}=\omega_{m_2}=\omega_m$ and $g_{0_1}=g_{0_2}=g_0$.

For further numerical computation, we work with re-scaled dynamical variables $\alpha_{1,2}=\left[\omega_m/\left(2\sqrt{\kappa_{s_{1,2}}}s^{\text{in}}_{1,2}\right)\right]a_{1,2}$, $\beta_{1,2}=\left(g_0/\omega_m\right)b_{1,2}$ and the normalized parameters with respect to $\omega_m$. Therefore, the equations of the motion are written as \cite{roque2020nonlinear,bakemeier2015route}
\begin{subequations}
    \begin{flalign}
    \frac{\text d\alpha_1}{\text d\tau}=&\left(i\frac{\Delta_1}{\omega_m}-\frac{\kappa_1}{2\omega_m}\right)\alpha_1-i2\text{Re}\left(\beta_1\right)\alpha_1
    \nonumber\\&+\frac{i\sqrt{\kappa_{e_1}}}{2\sqrt{\kappa_{s_1}}s^{\text{in}}_1}a^{\text{in}}_1+i\frac{1}{2},\\
    \frac{\text d\alpha_2}{\text d\tau}=&\left(i\frac{\Delta_2}{\omega_m}-\frac{\kappa_2}{2\omega_m}\right)\alpha_2-i2\text{Re}\left(\beta_2\right)\alpha_2\\\nonumber&+\frac{i\sqrt{\kappa_{e_2}}}{2\sqrt{\kappa_{s_2}}s^{\text{in}}_2}a^{\text{in}}_2+i\frac{1}{2},
    \\
    \frac{\text d\beta_{1,2}}{\text d\tau}=&-\left(i+\frac{\gamma_{m_{1,2}}}{2\omega_m}\right)\beta_{1,2}-i\frac{P_{1,2}}{2}|\alpha_{1,2}|^2,
\end{flalign}
\label{eq3}
\end{subequations}

\noindent where the dimensionless time is given by $\tau=\omega_mt$ and the dimensionless power is represented by $P_{1,2}=8\kappa_{s_{1,2}}(s^{\text{in}}_{1,2})^2 g_0^2/\omega_m^4$. Under our considered scenario, we set input field to cavity 1 $a^{\text{in}}_1(t)=0$ and the output field reduced to $a^{\text{out}}_1(t)=a^{\text{in}}_1(t)+i\sqrt{\kappa_{e_1}}a_1(t)=i\sqrt{\kappa_{e_1}}a_1(t)$. On the other hand, the input field $a^{\text{in}}_2$ of cavity 2 is the optical field coming from output field $a^{\text{out}}_1$ of cavity 1 which experiences time delay $t_d$ as well as accumulated phase $\phi_d$ due to propagation in the optical fiber; and additional adjustment of phase $\phi_c$ in the phase controller block. Therefore, the input field of the cavity 2 becomes $a^{\text{in}}_2(t)= \text e^{i\left(\phi_d+\phi_c\right)} a^{\text{out}}_1(t-t_d)$. Since the propagating distance is small, the effect of time delay $t_d$ can be safely ignored and only the net coupling phase, $\phi_d+\phi_c$ matters. Thereby, the coupling term in the slave cavity in Eq. (\ref{eq3}b) is written as
\begin{flalign}
    \frac{i\sqrt{\kappa_{e_2}}}{2\sqrt{\kappa_{s_2}}s^{\text{in}}_2}a^{\text{in}}_2(t)&=-\frac{\sqrt{\kappa_{e_1}\kappa_{e_2}}}{2\sqrt{\kappa_{s_2}}s^{\text{in}}_2} \text e^{i\left(\phi_d+\phi_c\right)}a_1(t)\nonumber\\
    \frac{i\sqrt{\kappa_{e_2}}}{2\sqrt{\kappa_{s_2}}s^{\text{in}}_2}a^{\text{in}}_2(\tau)&=-\sqrt{\frac{\kappa_{e_1}}{\omega_m}\frac{\kappa_{e_2}}{\omega_m}}\frac{\sqrt{\kappa_{s_1}}s^{\text{in}}_1}{\sqrt{\kappa_{s_2}}s^{\text{in}}_2}\text e^{\left(\phi_d+\phi_c\right)}\alpha_1(\tau)
    \label{eq4}
\end{flalign}

\subsection{Conditions for Chaos Synchronization}
In this subsection, the conditions for synchronization is derived. Complete chaos synchronization between the coupled optomechanical cavities is achievable when the mathematical form of the dynamical equations of the cavities in Eq. (\ref{eq3}) become identical. Now, we assume that the solution of Eq. (\ref{eq3}) lies on the synchronization manifold, that is $\boldsymbol{x_1}(\tau)=\boldsymbol{x}_2(\tau)\equiv \boldsymbol{s}(\tau)$ in which $\boldsymbol{x}_1=(\alpha_1,\beta_1)$, $\boldsymbol{x}_2=(\alpha_2,\beta_2)$ and $\boldsymbol{s}=(\alpha,\beta)$; and substituted them in Eqs. (\ref{eq3}). Thereby, we obtain the necessary conditions for complete synchronization (i.e., the existence of the synchronization manifold) in which the decay rates needs to be identical, which is,
\begin{equation}
    \kappa_1=\kappa_2\;\; \text{and}\;\; \gamma_{m_1}=\gamma_{m_2},
    \label{eq5}
\end{equation}
dimensionless power needs to equivalent, that is,
\begin{gather}
        P_1=P_2\; \text{or} \; \sqrt{\kappa_{s_1}}s^{\text{in}}_1=\sqrt{\kappa_{s_2}}s^{\text{in}}_2,
        \label{eq6}
\end{gather}
and lastly with the use of the condition in Eq. (\ref{eq6}) the detuning of the two cavities satisfy 
\begin{equation}
\Delta_1-\Delta_2=\sqrt{\kappa_{e_1}\kappa_{e_2}}\frac{\sqrt{\kappa_{s_1}}s^{\text{in}}_1}{\sqrt{\kappa_{s_2}}s^{\text{in}}_2}=\sqrt{\kappa_{e_1}\kappa_{e_2}},
\label{eq7}
\end{equation}
provided the coupling phase factor in Eq. (\ref{eq4}) is maintained at locked phase $\phi_{\text{lock}}$ such that
\begin{equation}
    \text e^{i\phi_{\text{lock}}}=-i.
    \label{eq8}
\end{equation}
where $\phi_{\text{lock}}=\phi_d+\phi_c$. The choice of imaginary coupling phase factor in Eq. (\ref{eq8}) ensures that detuning in cavity 2 effectively matches that of detuning in cavity 1 (that is, $\Delta_2^{\text{effective}}=\Delta_2+\sqrt{\kappa_{e_1}\kappa_{e_2}}=\Delta_1$) at the synchronization manifold. If the other mentioned parameters are equal, similar dynamical response is expected in the cavities and thus complete chaos synchronization is an allowed solution of Eqs. (\ref{eq3}). Since $\kappa_{e_{1,2}}\neq0$, the actual detuning in the cavities has to be different, which is $\Delta_1\neq\Delta_2$, and thereby follows a counter-intuitive approach as in general cases all the possible parameters in coupled optical nonlinear oscillators has to be equivalent. Here, the different detuning results in spectral separation of optical modes which can lead to suppression of the optical interaction between the cavities \cite{zhang2012synchronization}. However, as seen from Eq. (\ref{eq7}), the difference between $\omega_{c_1}$ and $\omega_{c_2}$ is within cavity linewidth $\kappa$ (since $\sqrt{\kappa_{e_1}\kappa_{e_2}}<\kappa$), which ensures that there is spectral overlap of the optical modes. There can be also alternative situation for complete synchronization, where 
\begin{equation}
    \text{e}^{i\phi_{\text{lock}}}=i,
    \label{eq9}
\end{equation}
leading to the condition
\begin{equation}
    \Delta_2-\Delta_1=\sqrt{\kappa_{e_1}\kappa_{e_2}},
    \label{eq10}
\end{equation}
but it induce poor synchronization which is further discussed later. The adjustment of the detuning values is feasible experimentally by changing the optical resonance frequency in each of the cavity \cite{koehler2018direct,zhang2012synchronization}. The propagation delay phase $\phi_d$ is determined by the length of the optical fiber, while the $\phi_c$ is adjusted by the phase controller block,  and therefore, the combination of these decides the overall coupling phase factor.

\begin{figure*}[t]
    \centering
    \includegraphics[width=0.85\linewidth]{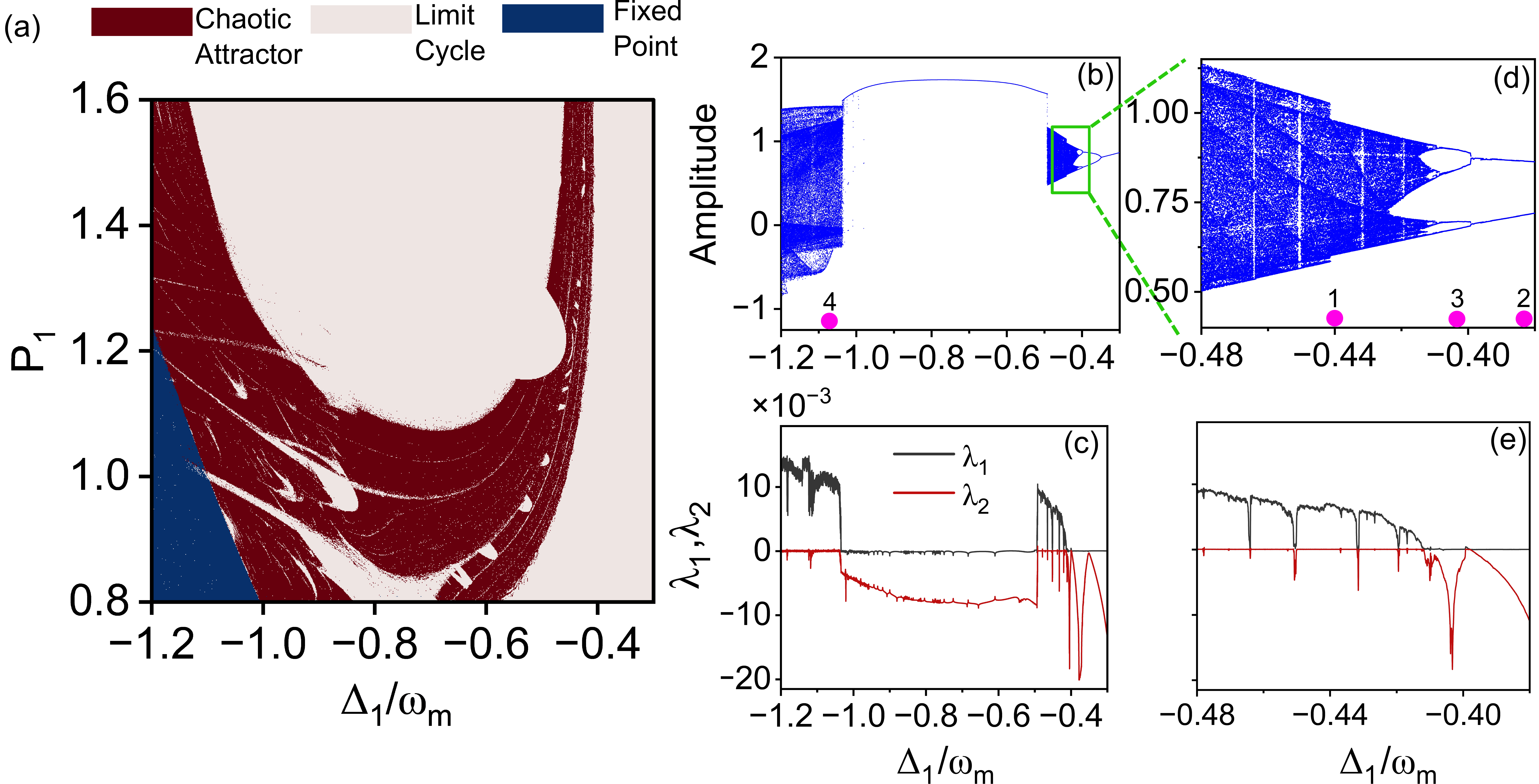}
    \caption{(a) The parameter space of $P_1$ and $\Delta_1$ of master cavity (cavity 1) illustrating regions of fixed point, limit cycle and chaotic attractor. The same is valid for cavity 2 as well, if uncoupled. (b) The bifurcation diagram of the mechanical oscillation amplitude under varying $\Delta_1$ at $P_1=1.3$. (c) The corresponding Lyapunov spectrum, consisting of two largest exponents ($\lambda_1$ and $\lambda_2$). (d)-(e) The magnified bifurcation diagram of the boxed part and it corresponding Lyapunov spectrum. The different dots labeled $1$, $2$, $3$ and $4$ are the considered operational points in the subsequent sections. }
    \label{fig2}
\end{figure*}
\subsection{Metrics to identify complete synchronization}
In our study, the quality of chaos synchronization is quantified in terms of the correlation coefficient involving optical intracavity intensity dynamics $I_{1,2}=|\alpha_{1,2}|^2$,
\begin{equation}
   \mathcal{C}=\frac{\langle [I_1(\tau)-\langle I_1(\tau)\rangle]\left[I_2(\tau)-\langle I_2(\tau)\rangle\right]\rangle}{\sqrt{\langle [I_1(\tau)-\langle I_1(\tau)\rangle]^2\rangle\langle[I_2(\tau)-\langle I_2(\tau)\rangle]^2\rangle}} ,
   \label{eq11}
\end{equation}
where $\langle\cdot\rangle$ is the time average of intracavity intensity dynamics. 

In addition, the stability of the complete synchronization manifold is addressed by calculating the maximum of transverse Lyapunov exponent ($\lambda_\perp^{\max}$) obtained from the variational equation of error dynamics \cite{gauthier1996intermittent}. For stability, \(\lambda_\perp^{\max}\) must have negative value, indicating that any deviation from the synchronization manifold will eventually revert back to the manifold. The error between the cavities is given by $\boldsymbol{e}(\tau)=\boldsymbol{x_1}(\tau)-\boldsymbol{x_2}(\tau)$ and here the vectors $\boldsymbol{x_1},\boldsymbol{x_2}$ are reassigned to real elements by expressing $\alpha_{1,2}=\alpha_{r_{1,2}}+i\alpha_{i_{1,2}}$ and $\beta_{1,2}=\beta_{r_{1,2}}+i\beta_{i_{1,2}}$. The vectors of small quantities are then written as $\boldsymbol{\delta x_{1,2}}=(\delta\alpha_{r_{1,2}},\delta\alpha_{i_{1,2}},\delta\beta_{r_{1,2}},\delta\beta_{i_{1,2}})$ and therefore $\boldsymbol{\delta e}=\boldsymbol{\delta x_1}-\boldsymbol{\delta x_2}$. The linearized equation of the error dynamics $\boldsymbol{\delta e_{1,2}}$ is written as
\begin{equation}
    \frac{\text d\boldsymbol{\delta e}}{\text d\tau}=\boldsymbol{M}\boldsymbol{\delta e},
    \label{eq12}
\end{equation}
where $\boldsymbol{M}$ is the Jacobian matrix calculated at the synchronization manifold $\boldsymbol{s}=(\alpha_r,\alpha_i,\beta_r,\beta_i)$ given by
\begin{equation}
\boldsymbol{M}=
    \begin{pmatrix}
        -\frac{\kappa}{2\omega_m} & -\frac{\Delta_2}{\omega_m}+2\beta_r & 2\alpha_i & 0\\
        \frac{\Delta_2}{\omega_m}-2\beta_r & -\frac{\kappa}{2\omega_m} & -2\alpha_r & 0\\
        0 & 0 & -\frac{\gamma_m}{2\omega_m} & 1\\
        -P\alpha_r & -P\alpha_i & -1 & -\frac{\gamma_m}{2\omega_m}\\
    \end{pmatrix}
    ,
    \label{eq13}
\end{equation}
and $\Delta_2=\Delta_1-\sqrt{\kappa_{e_1}\kappa_{e_2}}$, satisfying Eq. (\ref{eq7}). Note that the elements of $\boldsymbol{s}$ are reassigned to have real quantities as well.

The phenomenon of chaos synchronization can also be understood in terms of upper bound estimations for the mutual information rate (MIR) and Kolmogorov-Sinai (KS) entropy of the coupled system. KS entropy quantifies the information generated due to the uncertainty in the measurement of the chaotic trajectory. The upper bound of it, represented by $H_{\text{KS}}$ is given by the summation of positive Lyapunov exponents \cite{pesin1977characteristic}, $H_{\text{KS}}=\sum_j\lambda_j$, where $j$ is the index denoting positive Lyapunov exponents and the exponents are arranged in the order, $\lambda_1\geq\lambda_2\geq\cdots\geq\lambda_8$. The Lyapunov exponents of a dynamical system measures the rate of divergence of infinitesimally close trajectories in the phase space \cite{LU20051879,ott1981strange}. MIR quantifies the exchange of information between two variables, time series or dynamical systems. The upper bound of MIR, represented by $I_C$, can be estimated in terms of the two largest Lyapunov exponents of the coupled dynamical system \cite{baptista2012mutual},
\begin{equation}
    I_C=\lambda_1-\lambda_2.
    \label{eq14}
\end{equation}
For periodic orbits, the largest Lyapunov exponent is zero and the second is negative. Strictly speaking, substituting into Eq. (\ref{eq14}) would yield a positive number, but since periodic dynamics do not generate information, the MIR is conventionally taken as zero. The effect of chaos synchronization on both of the metrics is discussed in the subsequent sections.

\section{Dynamical regimes in single OMC}\label{sec3}
In this section, the numerical results are obtained for single uncoupeld OMC. The nonlinear regimes are analysed in the parameter space of dimensionless power ($P_1$ or $P_2$) and detuning ($\Delta_1$ or $\Delta_2$). Equations (\ref{eq3}) are numerically solved for $\kappa_{e_{1,2}}=0$, with the parameters $\kappa_1$ and $\gamma_{m_1}$ fixed at $0.73\omega_m$ and $7.7\times10^{-3}\omega_m$ respectively \cite{monifi2016optomechanically}, while $P_1$ and $\Delta_{1}$ are varied. The initial states of the optical dynamical variables are zeros since the photons lie in the ``cold" state and the initial state of the mechanical dynamical variables considered to possess zero displacement. We identified fixed point attractor regime in the large values of detuning in Fig. \ref{fig2}(a). The nonlinear regimes are the period-$n$ limit cycles and chaos. The chaotic regimes are confirmed by the positive maximum Lyapunov exponent. The chaotic regime at lower $P_1$ values in Fig. \ref{fig2}(a) spreads across wide range of $\Delta_1$, with self-similar periodic extended structures appearing all over where chaos is found. These structures known as shrimps \cite{gallas1993structure,maranhao2008experimental,medeiros2010periodic,de2024quasiperiodic} have similar shape of one, and appear to be aligned as if they were in a shrimp kabob stick. From about $P_1\approx1.07$, the chaotic regime occupies a curved region resembling open arms; one in the low detuned values, while other in the high detuned values. As $P_1$ increases, the chaotic regime in the left arm starts to diverge further from the right arm, with periodic limit cycles occupying the region between them and the regime in the right arm gradually reduces to non-chaotic state. This curved shape of the chaotic-periodic region is an indication of that the shrimps are occupying via homoclinic bifurcations scenario \cite{medrano2010periodic}. 

Now, for a fixed $P_1$, say at $P_1=1.3$, a bifurcation diagram of the peaks of the mechanical oscillation amplitude is plotted in Fig. \ref{fig2}(b) by sweeping the detuning parameter. The system goes from periodic to chaotic state through period doubling bifurcations as the detuning is increased. At $\Delta_1\approx-0.49\omega_m$, the system abruptly goes to self-induced oscillations (period-1 limit cycle) and then again abruptly switches to chaotic state at $\Delta_1\approx-1.04\omega_m$. This being the results of a crisis \cite{grebogi1983crises}. The corresponding plot of the Lyapunov spectrum of two largest exponents ($\lambda_1>\lambda_2$) is shown in Fig. \ref{fig2}(c) where the chaotic state corresponds to positive $\lambda_1$. To observe more of the features of chaotic regime a magnified plot of the boxed part in Fig. \ref{fig2}(b), is shown in Fig. \ref{fig2}(d) where multiple narrow windows of limit cycle are observed. The Lyapunov spectrum in Fig. \ref{fig2}(e) also confirms the same as $\lambda_1\approx0$ corresponding to those narrow windows.

\section{Process of complete chaos synchronization} \label{sec4}
\begin{figure}
    \centering
    \includegraphics[width=\linewidth]{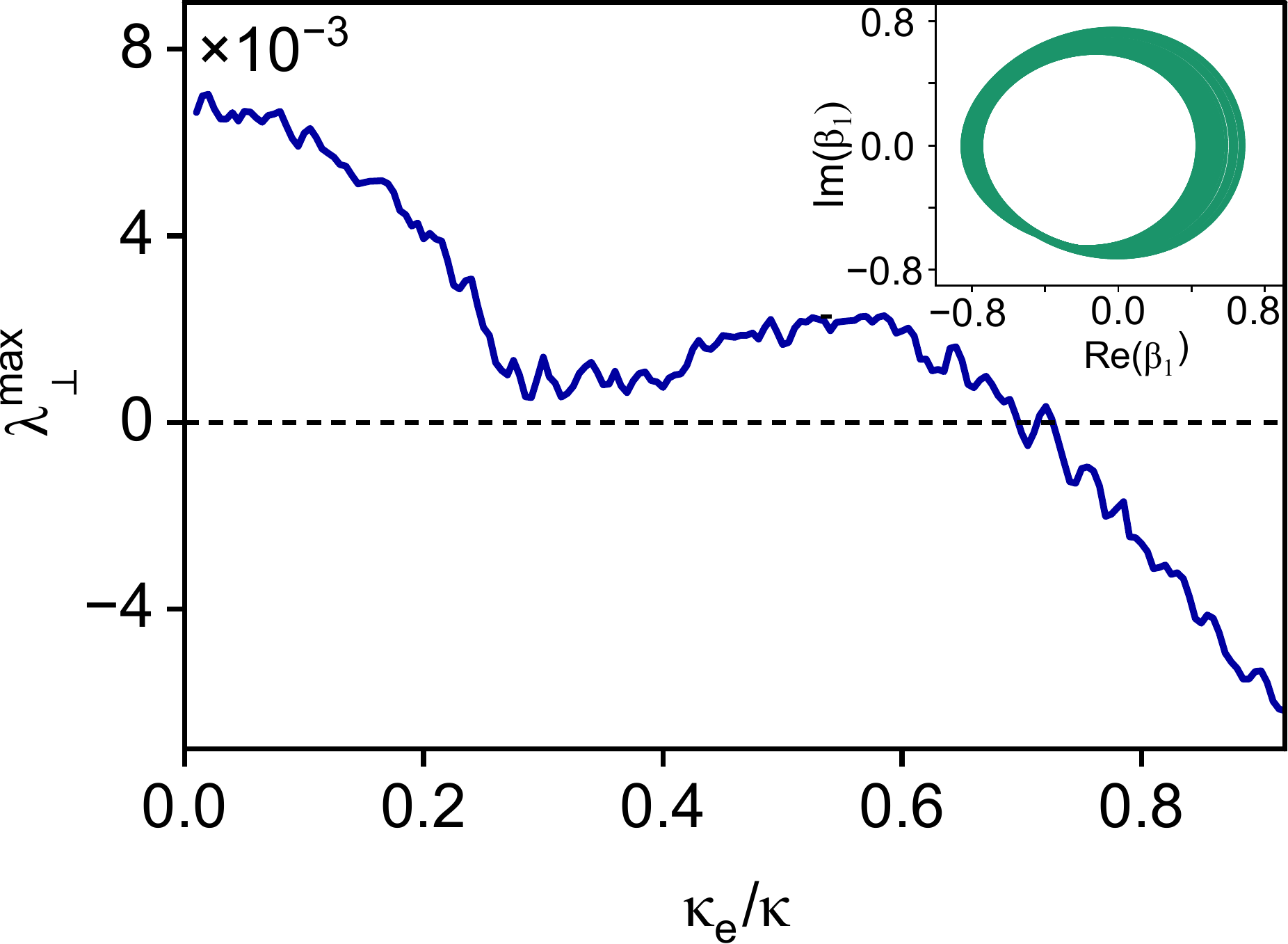}
    \caption{ The variation of the largest transverse Lyapunov exponent under varying external coupling rate $\kappa_e$,in which the stable region appears at $\kappa_e\gtrsim0.69\kappa$. The inset shows the chaotic attractor at the master cavity in the mechanical phase plane.}
    \label{fig3}
\end{figure}
\begin{figure}
    \includegraphics[width=\linewidth]{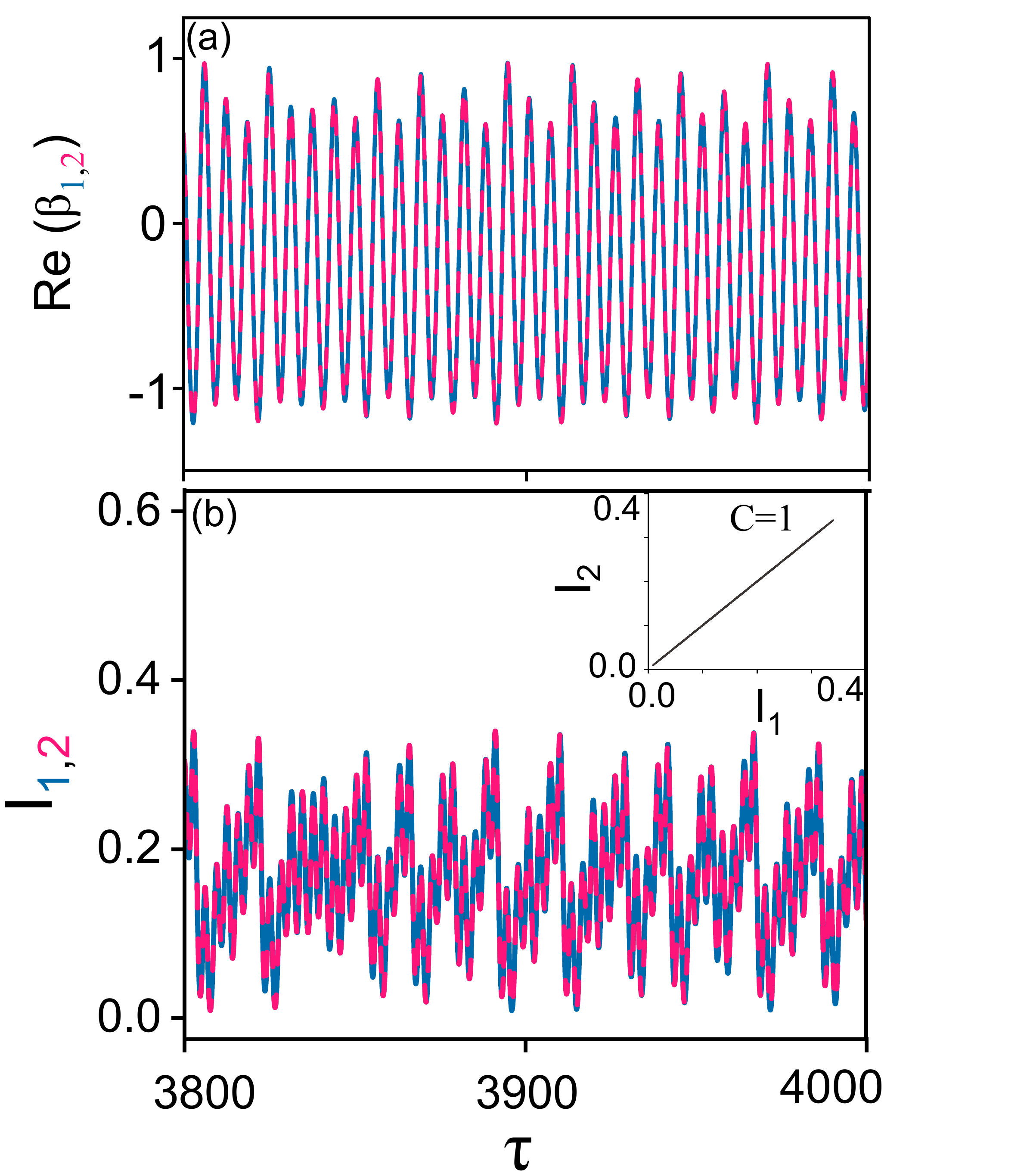}
    \caption{(a) The identical chaotic time series of mechanical oscillations of master and slave cavity. (b) The corresponding intracavity dynamics that are completely synchronized as well. The inset shows a projection of the chaotic attractor in the $I_1$-$I_2$ plane with perfect correlation of $C=1$ between the two time series. Note that, $\Delta_1=-0.44\omega_m$ and $\kappa_e=0.9\kappa$; and thereby $\Delta_2=\Delta_1-\kappa_e=-1.097\omega_m$ }
    \label{fig4}
\end{figure}

The cavities are now coupled, i.e., $\kappa_{e_{1,2}}\neq0$ and thereby, the phenomenon of chaos synchronization is studied, going from zero coupling value to substantial amount of coupling between the cavities. In the numerical simulations, the cavities are operated at same power level satisfying Eq. (\ref{eq6}), which is $P_1=P_2=P=1.3$. The decay rates are also kept identical satisfying Eq. (\ref{eq5}), that is, $\kappa_1=\kappa_2=\kappa=0.73\omega_m$, $\gamma_{m_1}=\gamma_{m_2}=\gamma_m=7.7\times10^{-3}\omega_m$. For convenience, we considered external coupling rate to be equal , which is, $\kappa_{e_1}=\kappa_{e_2}=\kappa_e$. We also maintained the condition in Eq. (\ref{eq8}). The detuning in the master cavity is chosen at $\Delta_1=-0.44\omega_m$ (dot labeled $1$ in Fig. \ref{fig2}). The stability of the synchronization manifold is assessed by examining the largest transverse Lyapunov exponents, as the external coupling rate $\kappa_e$ is varied from low to high magnitude. While varying $\kappa_e$, the detuning in the slave is varied according to Eq. (\ref{eq7}) such that the condition for complete synchronization remains valid. The plot in Fig. \ref{fig3} suggests that the synchronization manifold is unstable in the lower coupling value since $\lambda^{\text{max}}_\perp$ is positive. But as the coupling increases towards high magnitude (here, high means that the external coupling rate dominates the total cavity decay rate), $\lambda^{\text{max}}_\perp$ becomes negative (starting from $\kappa_e\approx0.69\kappa$) which indicates that the synchronization state is stable. A typical identical chaotic time series of master and slave cavity is plotted in Fig. \ref{fig4}, corresponding to the stable region at $\kappa_e=0.9\kappa$. Figure \ref{fig4}(a) shows the mechanical oscillation time series and Fig. \ref{fig4}(b) is the corresponding intra-cavity intensity dynamics, $I_{1,2}=|\alpha_{1,2}|^2$. The inset of Fig. \ref{fig4}(b) displays the projection of the chaotic attractor of the coupled system in the $I_1$-$I_2$ plane, which shows a identical line. This implies that the chaotic trajectory of the slave cavity lies in the synchronization manifold in a stable manner.  
\begin{figure}
    \centering
    \includegraphics[width=\linewidth]{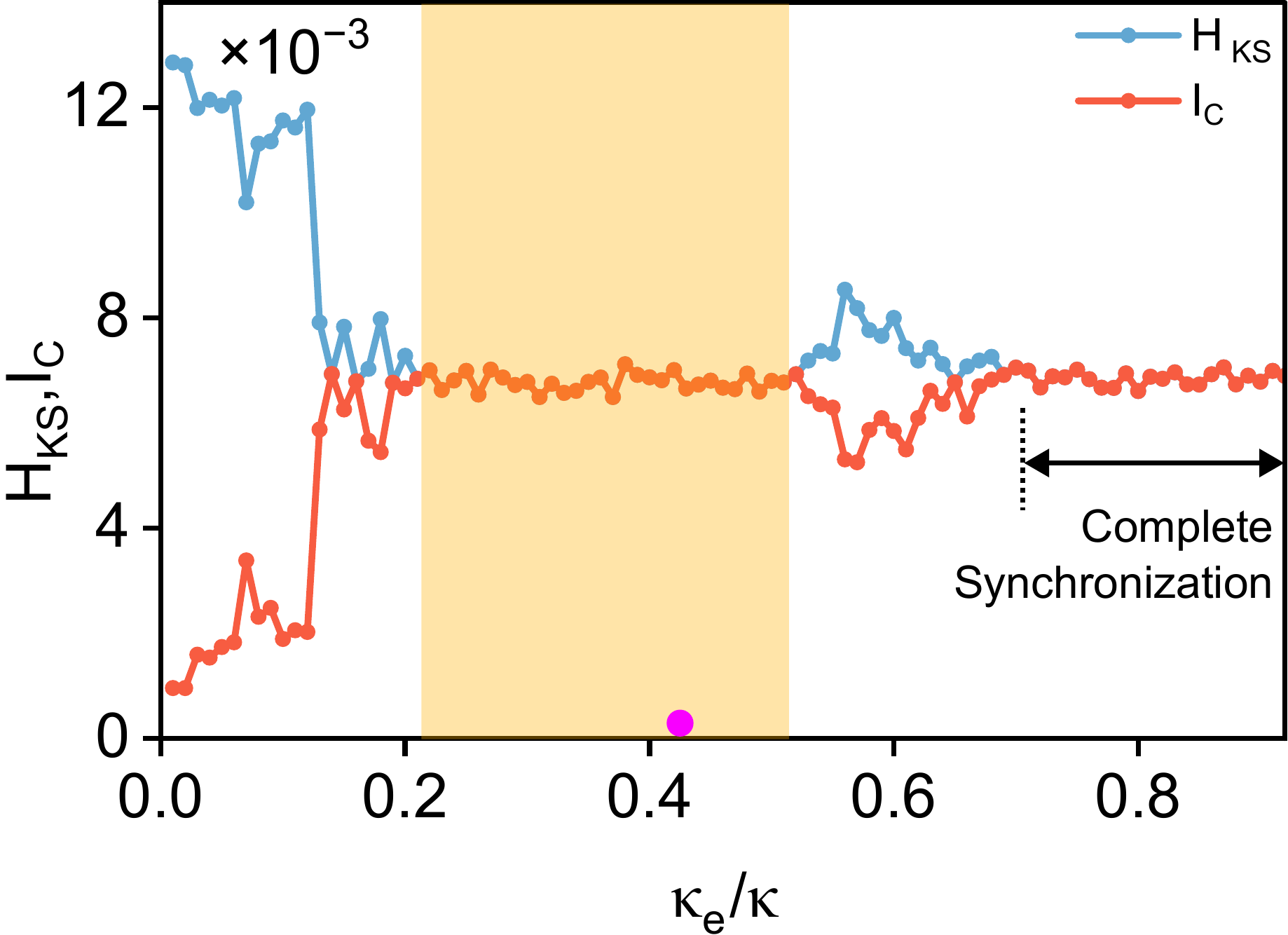}
    \caption{$H_{\text{KS}}$ and $I_C$ under varying external coupling rate $\kappa_e$. The parameters are same as in Fig. \ref{fig4} expect for $\Delta_2$ and $\kappa_e$. The units of these metrics are in bits per time unit. }
    \label{fig5}
\end{figure}
\begin{figure}
    \centering
    \includegraphics[width=\linewidth]{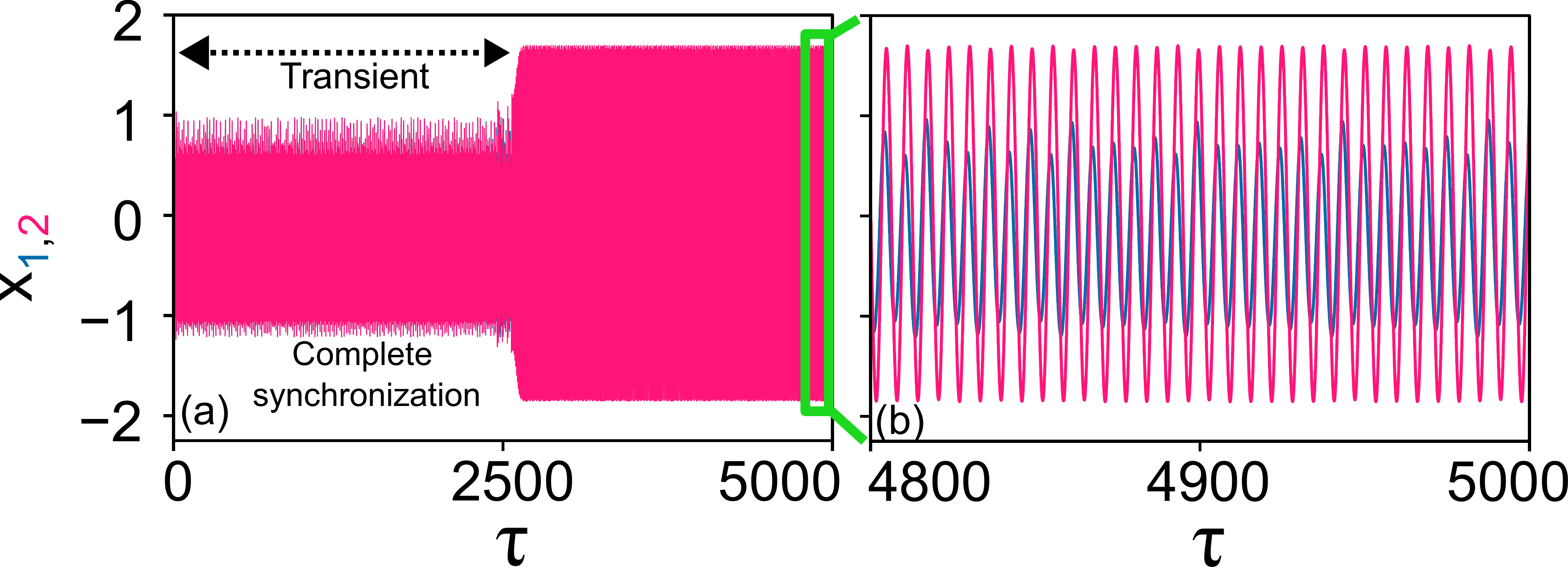}
    \caption{The mechanical oscillations of both master and slave at $\kappa_e=0.42\kappa$, that corresponds to the indicated dot in Fig. \ref{fig5}. (b) A magnified portion of the dynamics showing the asymptotic response of slave becoming different than that of master cavity.}
    \label{fig6}
\end{figure}
\begin{figure}
    \centering
    \includegraphics[width=\linewidth]{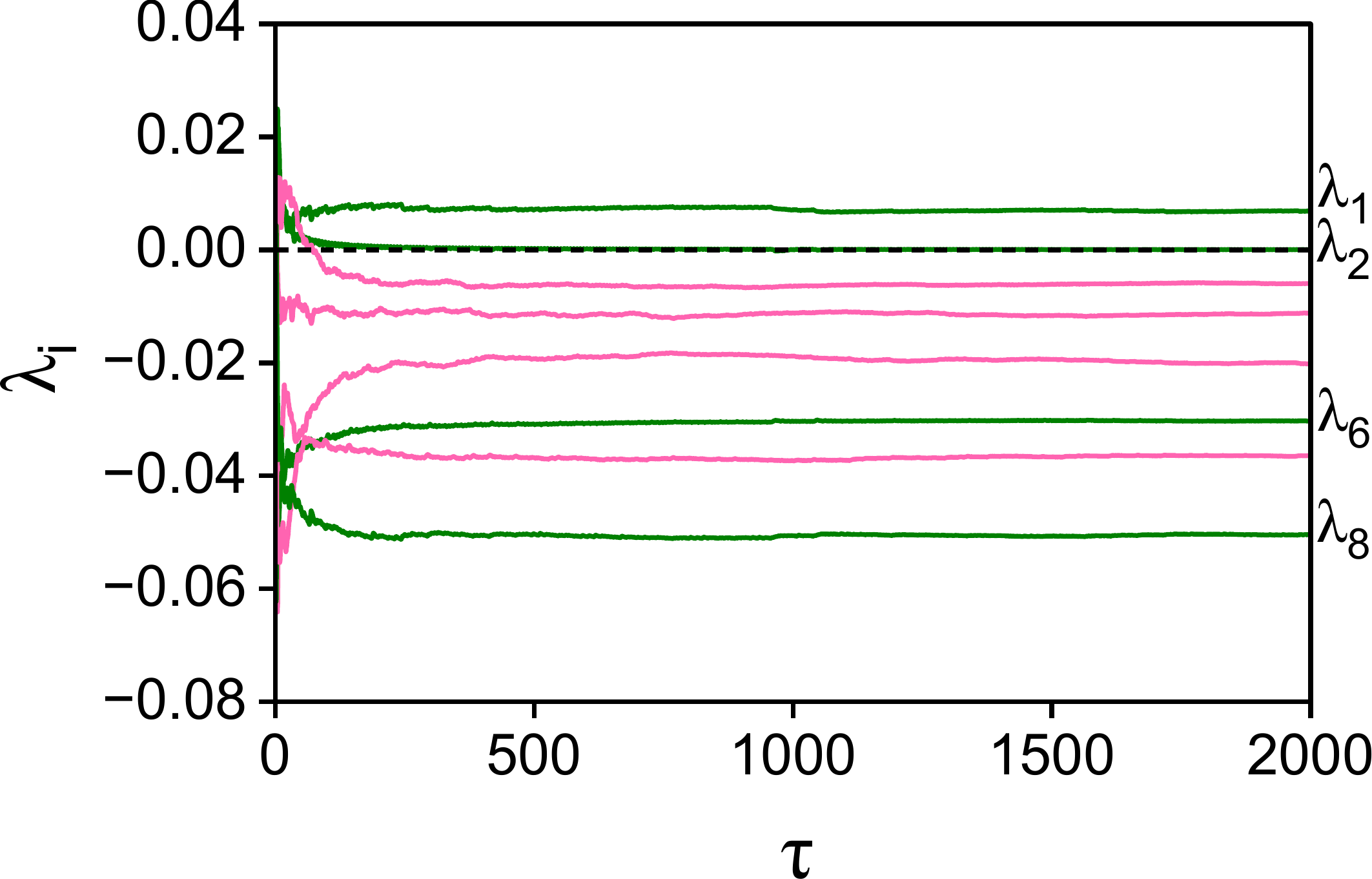}
    \caption{The spectrum of Lyapunov exponents $\lambda_i$ of the whole coupled dynamical equations during the stable synchronization condition. The parameters are same as in Fig. \ref{fig4}}
    \label{fig7}
\end{figure}
\begin{figure}
    \centering
    \includegraphics[width=\linewidth]{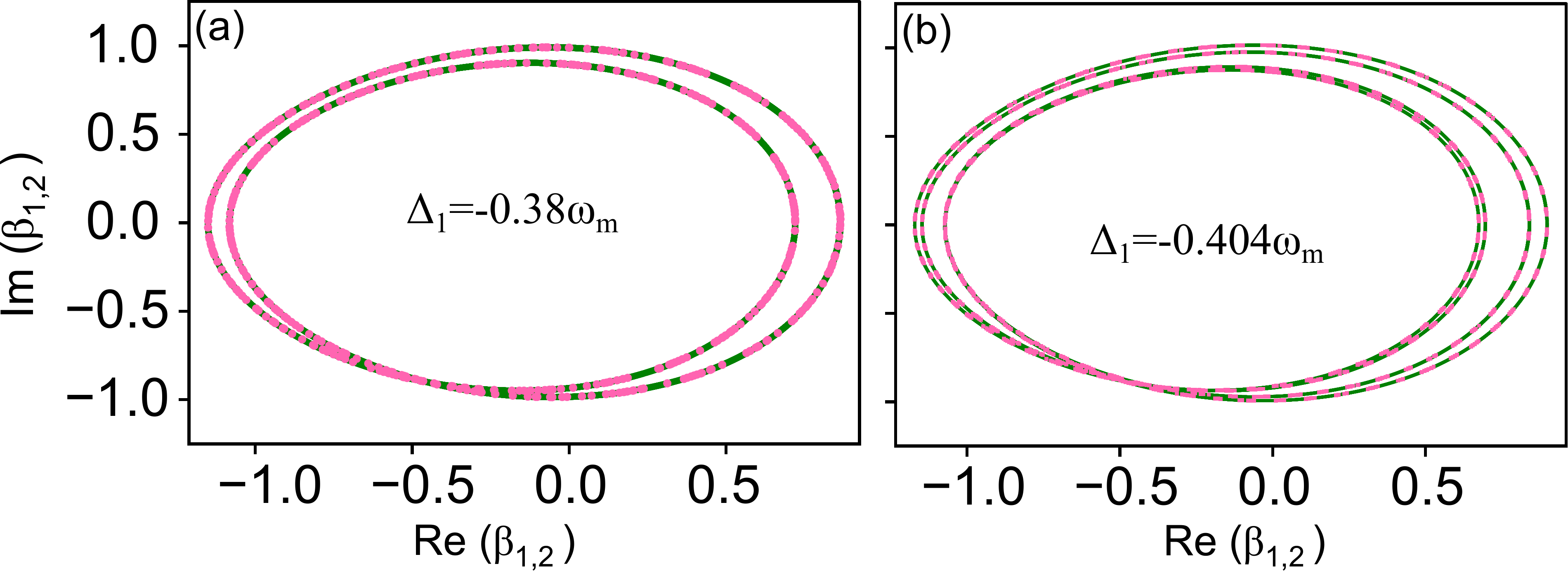}
    \caption{The trajectory of the slave cavity marked with dots in the mechanical phase space for (a) period-2 limit cycle and (b) period-4 limit cycle in the master cavity, which are marked by solid lines. The trajectory of the slave overlap with that of the master cavity, implying complete synchronization.}
    \label{fig8}
\end{figure}

Further understanding of the route to stable chaotic synchronization with respect to changing of coupling strength, is gained through $H_{\text{KS}}$ and $I_C$. During stable complete synchronization, the chaotic trajectory of the slave lies in the synchronization manifold and there is no information leaking in the transverse direction of the manifold (implying $\lambda_2=0$). Correspondingly, $H_{\text{KS}}=\lambda_1$ and also $I_C=\lambda_1$ from the definition, which makes $H_{\text{KS}}=I_C$. This is a route to synchronization which was studied in detail in \cite{baptista2005chaotic,baptista2008transmission}. In our system, the slave gets to know about the information of master through information exchange in the optical fiber during complete synchronization. From Fig. \ref{fig5}, we notice that at low coupling rate, the $H_{\text{KS}}$ of the coupled system is high, as both the cavities are nearly decoupled and thus it is possible to consider that the total entropy is close to the sum of the separated entropies of each system. Also, the information exchange between the two cavities is low as evident from low values of $I_c$, this being a consequence of the fact that the cavities are nearly decorrelated, and desynchronous. As the coupling is increased, more information is available at the slave cavity as evidenced by the increase of $I_C$ and thereby, the upper bound for the entropy of the system decreases as well, a consequence of the fact that master and slave become more synchronous. Further increase of coupling will lead the system into the shaded region in Fig. \ref{fig5} where $H_{\text{KS}}$ fluctuates around a certain value and also $I_C=H_{\text{KS}}$. Only based on the this equality, this parameter region can be mistaken to one leading the systems to complete chaos synchronization, but in reality, this is a consequence of the fact that the coupled system has a single positive Lyapunov exponent, and thus, by definition $H_{\text{KS}}=I_C$. However, the synchronization manifold is unstable, which is evident from the stability analysis in Fig. \ref{fig3}, and master and slave are desynchronous. This is also reaffirmed by the temporal mechanical dynamics in Fig. \ref{fig6} , at $\kappa_e=0.42\kappa$ lying in the shaded region in Fig. \ref{fig5}, which shows that the cavities are only able to sustain synchronized dynamics for a certain transient time window. Upon increasing the coupling, the system comes out of shaded region and there is increase of $H_{\text{KS}}$ as a consequence of the fact that a second Lyapunov exponent becomes positive, the system is higher-dimensional. Eventually, complete chaos synchronization starts from about $\kappa_e=0.69\kappa$ and $I_C=H_{\text{KS}}$. The value of the threshold coupling rate for complete synchronization comes out to be similar to that of stability analysis in Fig. \ref{fig3}. Now the plot in Fig. \ref{fig7} shows the Lyapunov spectrum of the coupled system, where the values of $\lambda_1$, $\lambda_2$, $\lambda_6$ and $\lambda_8$ correspond to the Lyapunov exponents of the chaotic attractor of the master cavity (easily verifiable by considering only the master cavity dynamical equation). As the trajectory lies on the synchronization manifold, these exponents are also equal to the conditional Lyapunov exponents along the direction of the synchronization manifold. And the remaining exponents are the transverse Lyapunov exponents along the transverse directions, the ones from which we can analyse the stability of the synchronization manifold (confirmed by obtaining the transverse Lyapunov spectrum from Eq. (\ref{eq12})-(\ref{eq13})). All the transversal Lyapunov exponents are negative, and thus this synchronization is stable.
\begin{figure}
    \centering
    \includegraphics[width=\linewidth]{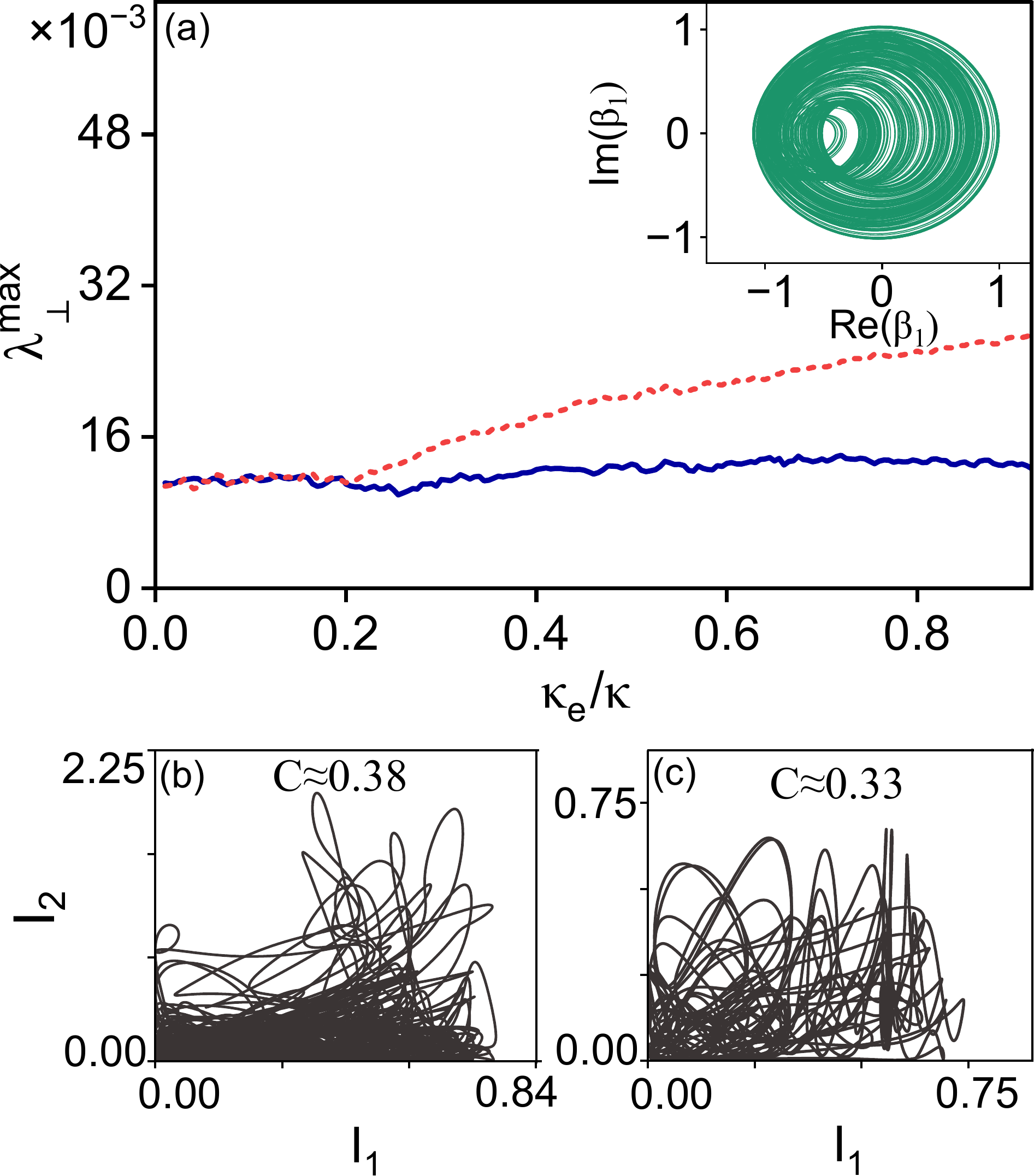}
    \caption{(a)The stability analysis of the synchronization manifold for the chaotic attractor (shown in the inset) with the master  at $\Delta_1=-1.097\omega_m$  under varying external coupling rate $\kappa_e$. The dashed curve corresponds to the case where alternate synchronization criteria is satisfied as given in Eq. (\ref{eq9})-(\ref{eq10}). (b) The projection of chaotic attractor of the coupled system in the $I_1$-$I_2$ plane at fixed $\kappa_e=0.9\kappa$. (c) The projection of chaotic attractor of the coupled system in the $I_1$-$I_2$ plane at fixed $\kappa_e=0.9\kappa$ for the alternate criterion of synchronization. }
    \label{fig9}
\end{figure}
\begin{figure*}
    \centering
    \includegraphics[width=\linewidth]{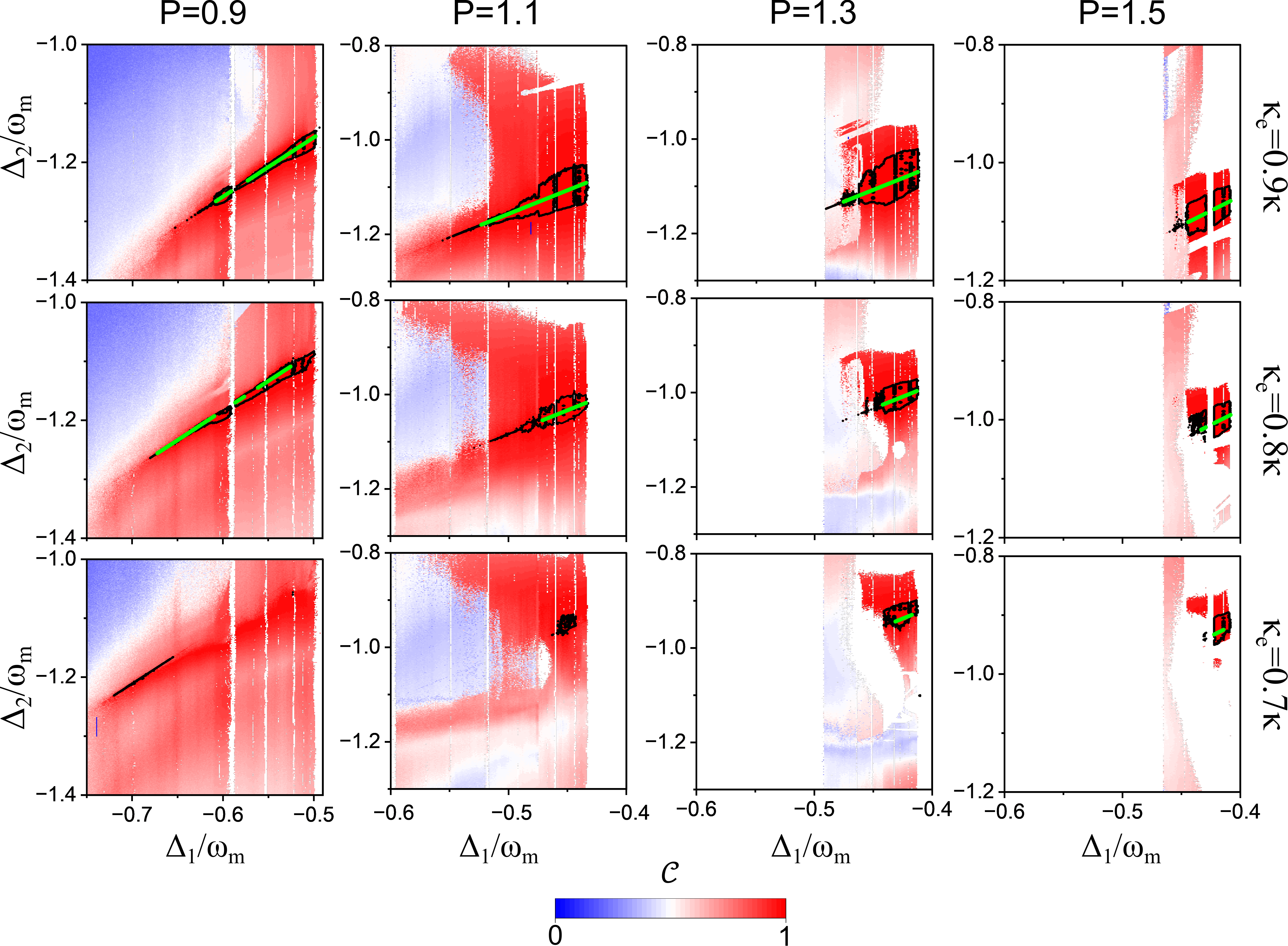}
    \caption{A detailed parameter space of the chaos synchronization quantified with correlation coefficient $C$ in the space of ($\Delta_1,\Delta_2$) for external coupling rate $\kappa_e= (0.7\kappa, 0.8\kappa, 0.9\kappa)$ and this is evaluated for $P = 0.9$, $P = 1.1$, $P = 1.3$ and $P = 1.5$. The green lines are the stable complete synchronization lines while black lines are the contour of $\mathcal{C}=0.99$. The white regions denote parameters that lead the master cavity to present periodic attractors }
    \label{fig10}
\end{figure*}

Till now, the analysis are made for the chaotic attractor of the master cavity at $\Delta_1=-0.44\omega_m$, but it is interesting to observe the behavior of the slave for  the master presenting other behaviours, such as periodic. We consider two periodic attractors for the master cavity, one is a period-2 limit cycle and another is a period-4 limit cycle occurring at $\Delta_1=-0.38\omega_m$ and $\Delta_1=-0.404\omega_m$ (dot labeled $2$ and $3$ respectively in Fig. \ref{fig2}), respectively. The detuning of the slave is tuned according to Eq. (\ref{eq7}) and the other parameters are same as in Fig. \ref{fig4}. The trajectory in the mechanical phase space in Fig. \ref{fig8} shows that the slave exactly follows the master cavity trajectory. Thereby, the synchronization condition is consistent for the master cavity with periodic attractors and we also say that slave cavity becomes completely synchronous to the master for each attractor within parameters leading to chaos via periodic doubling route.

Now, the master cavity exhibits chaotic attractor at different parameters, say at $\Delta_1=-1.097\omega_m$ (dot labeled $4$ in Fig. \ref{fig2}) where the largest Lyapunov exponent is greater compared to the chaotic attractor at $\Delta_1=-0.44\omega_m$. This indicates that the chaotic attractor is more sensitive to initial conditions. The trajectory in the mechanical phase space displays more complicated trajectory as shown in the inset of Fig. \ref{fig9}(a) and also occupies more region in the phase space. Now, a stability analysis is performed to find out the stable regions for complete synchronization. The external coupling rate is varied from low to high, and correspondingly $\Delta_2$ is tuned according to Eq. (\ref{eq7}) to keep the synchronization condition intact. The solid curve in Fig. \ref{fig9}(a) shows that the system is unable to stabilize for any values of the coupling rate. The projection of the chaotic attractor in Fig. \ref{fig9}(b) for $\kappa_e=0.9\kappa$ in the $I_1$-$I_2$ plane shows a complicated structure implying poor correlation $\mathcal{C}\approx0.38$.  The same analysis is performed for the case where the alternate criterion for synchronization given in Eq. (\ref{eq9}) and Eq. (\ref{eq10}), is satisfied. The dashed curve in Fig. \ref{fig9}(a) shows that this criterion also fails to stabilize the synchronization and in fact more harder to stabilize. Here too, the projection of the attractor in the $I_1$-$I_2$ plane in Fig. \ref{fig9}(c) displays a complicated relation with correlation of $\mathcal{C}\approx0.33$.  Overall, the  nature of the chaotic attractor in the master cavity plays an important role in stabilization. More chaotic the master cavity is, harder for the slave to achieve stability in synchronization.  

\section{Synchronization quality under parametric variation}\label{sec5}

In this section we discuss in detail how the quality of chaos synchronization behaves for different values of dimensionless power $P$ and locked phase $\phi_{\text{lock}}$. A two dimensional map, in the parameter space of ($\Delta_1$,$\Delta_2$) is plotted in Fig. \ref{fig10} depicting the quality of chaos synchronization in terms of coefficient $\mathcal{C}$ for four different values of power $P=(0.9,1.1,1.3,1.5)$ each evaluated at $\kappa_e=(0.7\kappa,0.8\kappa,0.9\kappa)$. The green lines in the plots indicate parameters for which complete synchronization of chaotic trajectories in master and in the slave is stable. By considering $C\geq0.99$ as the high quality synchronization region, we see that the quality is high around the green line (observe the black contour around this line), but decreases as we go away from it. For every values of $P$'s in the Fig. \ref{fig10}, the region of chaos synchronization with high correlation decreases when $\kappa_e$ is reduced. This is expected as the strength of the optical field in the fiber reduces and thereby influence of the master cavity on the slave becomes less as well, which reduces the correlation. High quality chaos synchronization still persists at lower $\kappa_e$ at higher $P$ (for example, follow the plots corresponding to $\kappa_e=0.7\kappa$) since higher $P$ compensate the reduced external coupling strength. On the other hand, the extent of stable complete synchronization line in the plots, corresponding to, for say $\kappa_e=0.9\kappa$, is more for $P=0.9$ and $P=1.1$, which is simply because of the accessibility of wider chaotic regions in those power levels.

\begin{figure}
    \centering
    \includegraphics[width=\linewidth]{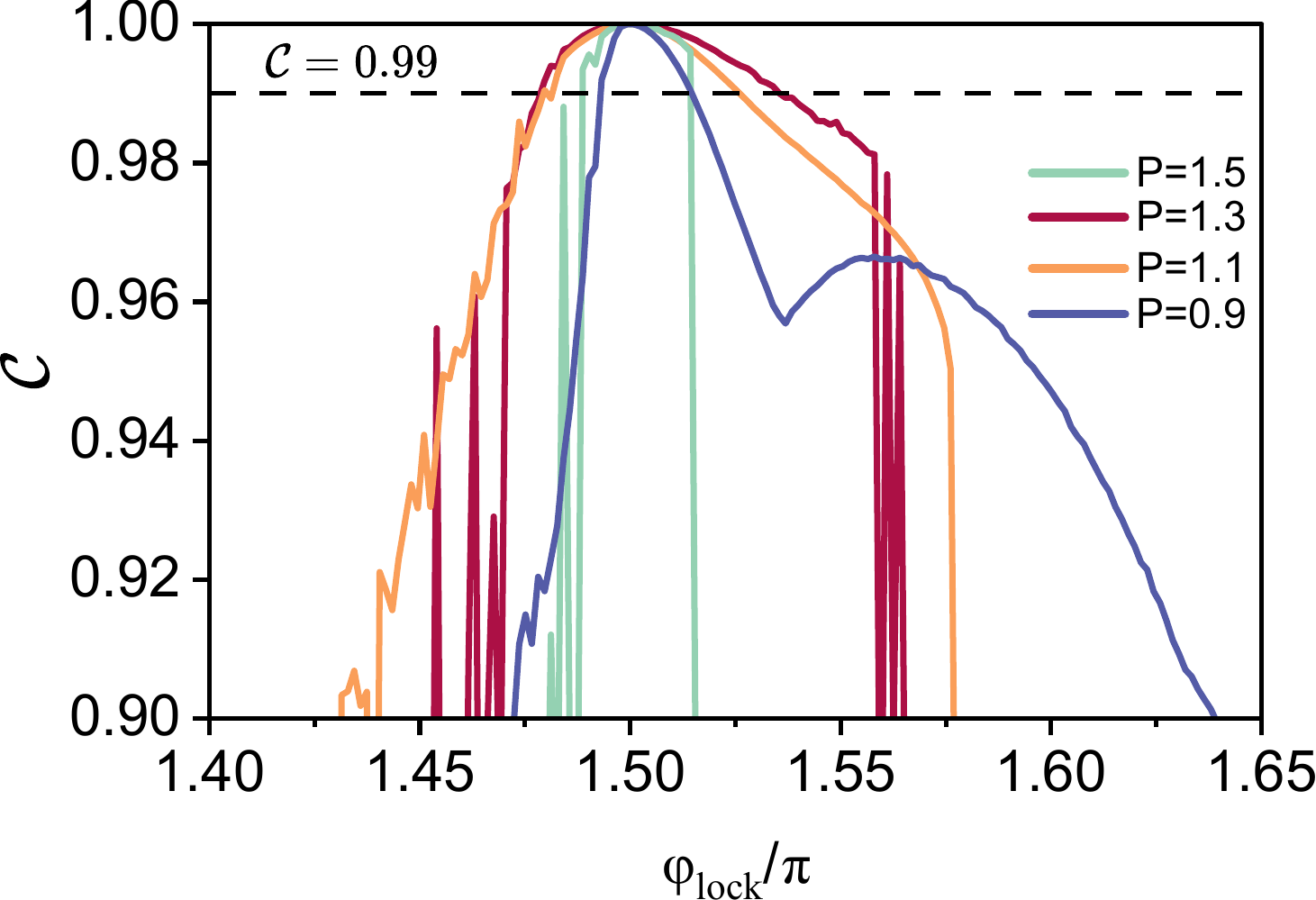}
    \caption{The changes in correlation strength against variation of locked phase for different values of $P$.}
    \label{fig11}
\end{figure}

Now, in an experimental setup, the locking of the phase $\phi_{\text{lock}}$ to desired value can be difficult and therefore, it is crucial to understand the robustness of synchronization against the variation of locked phase. Figure \ref{fig11} shows the plot of correlation strength against changes in the locked phase about $1.5\pi$ (here, this value is one of the solution of Eq. (\ref{eq8})). Increasing the power level to $P=1.5$ from $P=1.3$, reduces the robustness of the synchronization as the extent of the high $\mathcal{C}$ region decreases considerably. Similarly, decrease in the power levels to $P=0.9$ also reduces the extent of high $\mathcal{C}$ region. Therefore, operating the system between power levels $P=0.9$ and $P=1.5$, will provide relatively higher robustness against variation of locked phase. 

\section{Experimental Realization}\label{sec6}
The demonstration of complete chaos synchronization between optically mediated coupled optomechanical cavities is possible in future experimental demonstrations. The study shows that the total cavity decay rate is dominated by $\kappa_e$ term, which imply that intrinsic optical losses should be very low and thereby high quality factor (Q) cavity is required. Assuming the total optical decay rate $\kappa$ is in the order of $\omega_m$ and out of it, intrinsic loss rate $k_{i_{1,2}}\approx0.1\kappa$. By operating optical resonance frequency $\omega_{c_{1,2}}\sim2\pi\times10^{14}\text{Hz}$ and $\omega_m\sim2\pi\times10^7\text{Hz}$ \cite{monifi2016optomechanically}, the quality factor turns to be around $Q\sim10^8$. Recent progress in fabrication methods allows for realization of ultra high-Q ($\geq10^8$) optical microcavity \cite{honari2021fabrication,wu2020greater,gu2021dry}.
Operating with reduced $\kappa_e$ relaxes the requirement of high-Q cavity to a certain degree, but the availability of high correlation synchronization region reduces as evident from Fig. \ref{fig10}. Now, the input power $P_{\text{in}_{1,2}}$ for the cavities is given by the expression $s^{\text{in}}_{{1,2}}=\sqrt{P_{\text{in}_{1,2}}/\hbar\omega_L}$. Assuming $\omega_L=2\pi\times193\;\text{THz}$, $\omega_m=2\pi\times20\text{MHz}$, $g_0=10^{-4}\omega_m$, $\kappa=0.73\omega_m$, $\kappa_{s_{1,2}}=0.01\kappa$, the input power $P_{\text{in}_{1,2}}\approx24\;\text{mW}$, $P_{\text{in}_{1,2}}\approx35\;\text{mW}$ and $P_{\text{in}_{1,2}}\approx42\;\text{mW}$ for $P=0.9$, $P=1.3$ and $P=1.5$ respectively. Thereby we have an estimate of the magnitude of operable input power.
\section{Summary}\label{sec7}
In summary, we have demonstrated stable, complete chaos synchronization in a master-slave configuration of two optomechanical cavities coupled unidirectionally through an optical fiber. We derive the relationship among the parameters, where the difference between the detuning of the cavities should be equivalent to the external coupling rate for achieving complete synchronization. The stability analysis of the synchronization manifold shows that stable synchronization exists in the region of high magnitude of external coupling rate. The pathway to this stable synchronization begins with the system undergoing desynchronous chaotic responses in both the cavities at weak coupling. As the coupling increases, the system enters into low-dimensional state with unstable synchronization manifold by presenting a single positive Lyapunov exponent. Further increase of coupling makes the system high-dimensional with two positive exponents and eventually the system move towards stable complete synchronization regime, which is also characterized by single positive exponent. Our study indicates that the presence of a chaotic attractor in the master cavity, with a more complicated trajectory and higher sensitivity to initial conditions, de-stabilize the synchronization manifold which results in desynchronous temporal responses. Lastly, the study on the synchronization quality shows that strong correlated chaotic dynamics are accessible for a wide regime in the parameter space of the detunings of the cavities, which can provide increased flexibilty in experimental conditions. Overall, our work provided insight into the collective behaviour of OMCs in the chaotic regime and in periodic regime, which has the potential to extend to the case of a complex network. Since the cavities are optically coupled, current study could also be extended to understand long range interactions of the cavities where the optical transmission delay becomes prominent. The master-slave configuration shows potential for demonstrating long-distance secure optical communication provided that high mechanical resonance frequency OMCs are used and further feasibility studies regarding the impact of transmission delays as well as presence of complicated chaotic attractor in the master cavity, needs to be conducted.


\begin{thebibliography}{77}%
\makeatletter
\providecommand \@ifxundefined [1]{%
 \@ifx{#1\undefined}
}%
\providecommand \@ifnum [1]{%
 \ifnum #1\expandafter \@firstoftwo
 \else \expandafter \@secondoftwo
 \fi
}%
\providecommand \@ifx [1]{%
 \ifx #1\expandafter \@firstoftwo
 \else \expandafter \@secondoftwo
 \fi
}%
\providecommand \natexlab [1]{#1}%
\providecommand \enquote  [1]{``#1''}%
\providecommand \bibnamefont  [1]{#1}%
\providecommand \bibfnamefont [1]{#1}%
\providecommand \citenamefont [1]{#1}%
\providecommand \href@noop [0]{\@secondoftwo}%
\providecommand \href [0]{\begingroup \@sanitize@url \@href}%
\providecommand \@href[1]{\@@startlink{#1}\@@href}%
\providecommand \@@href[1]{\endgroup#1\@@endlink}%
\providecommand \@sanitize@url [0]{\catcode `\\12\catcode `\$12\catcode
  `\&12\catcode `\#12\catcode `\^12\catcode `\_12\catcode `\%12\relax}%
\providecommand \@@startlink[1]{}%
\providecommand \@@endlink[0]{}%
\providecommand \url  [0]{\begingroup\@sanitize@url \@url }%
\providecommand \@url [1]{\endgroup\@href {#1}{\urlprefix }}%
\providecommand \urlprefix  [0]{URL }%
\providecommand \Eprint [0]{\href }%
\providecommand \doibase [0]{https://doi.org/}%
\providecommand \selectlanguage [0]{\@gobble}%
\providecommand \bibinfo  [0]{\@secondoftwo}%
\providecommand \bibfield  [0]{\@secondoftwo}%
\providecommand \translation [1]{[#1]}%
\providecommand \BibitemOpen [0]{}%
\providecommand \bibitemStop [0]{}%
\providecommand \bibitemNoStop [0]{.\EOS\space}%
\providecommand \EOS [0]{\spacefactor3000\relax}%
\providecommand \BibitemShut  [1]{\csname bibitem#1\endcsname}%
\let\auto@bib@innerbib\@empty
\bibitem [{\citenamefont {Strogatz}(2018)}]{strogatz2018nonlinear}%
  \BibitemOpen
  \bibfield  {author} {\bibinfo {author} {\bibfnamefont {S.~H.}\ \bibnamefont
  {Strogatz}},\ }\href@noop {} {\emph {\bibinfo {title} {Nonlinear dynamics and
  chaos: With applications to physics, biology, chemistry, and engineering}}}\
  (\bibinfo  {publisher} {CRC press},\ \bibinfo {address} {Boca Raton},\
  \bibinfo {year} {2018})\BibitemShut {NoStop}%
\bibitem [{\citenamefont {Thiel}\ \emph {et~al.}(2010)\citenamefont {Thiel},
  \citenamefont {Kurths}, \citenamefont {Romano}, \citenamefont {Károlyi},\
  and\ \citenamefont {Moura}}]{marco2010}%
  \BibitemOpen
  \bibfield  {author} {\bibinfo {author} {\bibfnamefont {M.}~\bibnamefont
  {Thiel}}, \bibinfo {author} {\bibfnamefont {J.}~\bibnamefont {Kurths}},
  \bibinfo {author} {\bibfnamefont {M.~C.}\ \bibnamefont {Romano}}, \bibinfo
  {author} {\bibfnamefont {G.}~\bibnamefont {Károlyi}},\ and\ \bibinfo
  {author} {\bibfnamefont {A.}~\bibnamefont {Moura}},\ }\href@noop {} {\emph
  {\bibinfo {title} {Nonlinear Dynamics and Chaos: Advances and
  Perspectives}}}\ (\bibinfo  {publisher} {Springer},\ \bibinfo {address}
  {Berlin, Heidelberg},\ \bibinfo {year} {2010})\BibitemShut {NoStop}%
\bibitem [{\citenamefont {Ott}(1981)}]{ott1981strange}%
  \BibitemOpen
  \bibfield  {author} {\bibinfo {author} {\bibfnamefont {E.}~\bibnamefont
  {Ott}},\ }\bibfield  {title} {\bibinfo {title} {Strange attractors and
  chaotic motions of dynamical systems},\ }\href@noop {} {\bibfield  {journal}
  {\bibinfo  {journal} {Reviews of Modern Physics}\ }\textbf {\bibinfo {volume}
  {53}},\ \bibinfo {pages} {655} (\bibinfo {year} {1981})}\BibitemShut
  {NoStop}%
\bibitem [{\citenamefont {Pecora}\ and\ \citenamefont
  {Carroll}(1990)}]{PhysRevLett.64.821}%
  \BibitemOpen
  \bibfield  {author} {\bibinfo {author} {\bibfnamefont {L.~M.}\ \bibnamefont
  {Pecora}}\ and\ \bibinfo {author} {\bibfnamefont {T.~L.}\ \bibnamefont
  {Carroll}},\ }\bibfield  {title} {\bibinfo {title} {Synchronization in
  chaotic systems},\ }\href@noop {} {\bibfield  {journal} {\bibinfo  {journal}
  {Phys. Rev. Lett.}\ }\textbf {\bibinfo {volume} {64}},\ \bibinfo {pages}
  {821} (\bibinfo {year} {1990})}\BibitemShut {NoStop}%
\bibitem [{\citenamefont {Jovic}(2011)}]{jovic2011chaotic}%
  \BibitemOpen
  \bibfield  {author} {\bibinfo {author} {\bibfnamefont {B.}~\bibnamefont
  {Jovic}},\ }\bibfield  {title} {\bibinfo {title} {Chaotic signals and their
  use in secure communications},\ }in\ \href@noop {} {\emph {\bibinfo
  {booktitle} {Synchronization Techniques for Chaotic Communication Systems}}}\
  (\bibinfo  {publisher} {Springer},\ \bibinfo {address} {Berlin, Heidelberg},\
  \bibinfo {year} {2011})\ pp.\ \bibinfo {pages} {31--47}\BibitemShut {NoStop}%
\bibitem [{\citenamefont {Mishkovski}\ and\ \citenamefont
  {Kocarev}(2011)}]{mishkovski2011chaos}%
  \BibitemOpen
  \bibfield  {author} {\bibinfo {author} {\bibfnamefont {I.}~\bibnamefont
  {Mishkovski}}\ and\ \bibinfo {author} {\bibfnamefont {L.}~\bibnamefont
  {Kocarev}},\ }\bibfield  {title} {\bibinfo {title} {Chaos-based public-key
  cryptography},\ }in\ \href@noop {} {\emph {\bibinfo {booktitle} {Chaos-Based
  Cryptography: Theory, Algorithms and Applications}}},\ \bibinfo {editor}
  {edited by\ \bibinfo {editor} {\bibfnamefont {L.}~\bibnamefont {Kocarev}}\
  and\ \bibinfo {editor} {\bibfnamefont {S.}~\bibnamefont {Lian}}}\ (\bibinfo
  {publisher} {Springer},\ \bibinfo {address} {Berlin, Heidelberg},\ \bibinfo
  {year} {2011})\ pp.\ \bibinfo {pages} {27--65}\BibitemShut {NoStop}%
\bibitem [{\citenamefont {Zhang}\ \emph {et~al.}(2024)\citenamefont {Zhang},
  \citenamefont {Zhang}, \citenamefont {Qiao},\ and\ \citenamefont
  {Wang}}]{zhang2024chaos}%
  \BibitemOpen
  \bibfield  {author} {\bibinfo {author} {\bibfnamefont {M.}~\bibnamefont
  {Zhang}}, \bibinfo {author} {\bibfnamefont {J.}~\bibnamefont {Zhang}},
  \bibinfo {author} {\bibfnamefont {L.}~\bibnamefont {Qiao}},\ and\ \bibinfo
  {author} {\bibfnamefont {T.}~\bibnamefont {Wang}},\ }\bibfield  {title}
  {\bibinfo {title} {Chaos brillouin distributed optical fiber sensing},\ }in\
  \href@noop {} {\emph {\bibinfo {booktitle} {Novel Optical Fiber Sensing
  Technology and Systems}}}\ (\bibinfo  {publisher} {Springer},\ \bibinfo
  {address} {Singapore},\ \bibinfo {year} {2024})\ pp.\ \bibinfo {pages}
  {147--217}\BibitemShut {NoStop}%
\bibitem [{\citenamefont {Aihara}\ \emph {et~al.}(1990)\citenamefont {Aihara},
  \citenamefont {Takabe},\ and\ \citenamefont {Toyoda}}]{aihara1990chaotic}%
  \BibitemOpen
  \bibfield  {author} {\bibinfo {author} {\bibfnamefont {K.}~\bibnamefont
  {Aihara}}, \bibinfo {author} {\bibfnamefont {T.}~\bibnamefont {Takabe}},\
  and\ \bibinfo {author} {\bibfnamefont {M.}~\bibnamefont {Toyoda}},\
  }\bibfield  {title} {\bibinfo {title} {Chaotic neural networks},\ }\href@noop
  {} {\bibfield  {journal} {\bibinfo  {journal} {Physics letters A}\ }\textbf
  {\bibinfo {volume} {144}},\ \bibinfo {pages} {333} (\bibinfo {year}
  {1990})}\BibitemShut {NoStop}%
\bibitem [{\citenamefont {Soriano}\ \emph {et~al.}(2013)\citenamefont
  {Soriano}, \citenamefont {Garc{\'\i}a-Ojalvo}, \citenamefont {Mirasso},\ and\
  \citenamefont {Fischer}}]{soriano2013complex}%
  \BibitemOpen
  \bibfield  {author} {\bibinfo {author} {\bibfnamefont {M.~C.}\ \bibnamefont
  {Soriano}}, \bibinfo {author} {\bibfnamefont {J.}~\bibnamefont
  {Garc{\'\i}a-Ojalvo}}, \bibinfo {author} {\bibfnamefont {C.~R.}\ \bibnamefont
  {Mirasso}},\ and\ \bibinfo {author} {\bibfnamefont {I.}~\bibnamefont
  {Fischer}},\ }\bibfield  {title} {\bibinfo {title} {Complex photonics:
  Dynamics and applications of delay-coupled semiconductors lasers},\
  }\href@noop {} {\bibfield  {journal} {\bibinfo  {journal} {Reviews of Modern
  Physics}\ }\textbf {\bibinfo {volume} {85}},\ \bibinfo {pages} {421}
  (\bibinfo {year} {2013})}\BibitemShut {NoStop}%
\bibitem [{\citenamefont {Junji}(2017)}]{ohtsubo2017semiconductor}%
  \BibitemOpen
  \bibfield  {author} {\bibinfo {author} {\bibfnamefont {O.}~\bibnamefont
  {Junji}},\ }\href@noop {} {\emph {\bibinfo {title} {Semiconductor Lasers:
  Stability, Instability and Chaos}}}\ (\bibinfo  {publisher} {Springer},\
  \bibinfo {address} {Cham},\ \bibinfo {year} {2017})\BibitemShut {NoStop}%
\bibitem [{\citenamefont {Carmon}\ \emph {et~al.}(2007)\citenamefont {Carmon},
  \citenamefont {Cross},\ and\ \citenamefont {Vahala}}]{carmon2007chaotic}%
  \BibitemOpen
  \bibfield  {author} {\bibinfo {author} {\bibfnamefont {T.}~\bibnamefont
  {Carmon}}, \bibinfo {author} {\bibfnamefont {M.}~\bibnamefont {Cross}},\ and\
  \bibinfo {author} {\bibfnamefont {K.~J.}\ \bibnamefont {Vahala}},\ }\bibfield
   {title} {\bibinfo {title} {Chaotic quivering of micron-scaled on-chip
  resonators excited by centrifugal optical pressure},\ }\href@noop {}
  {\bibfield  {journal} {\bibinfo  {journal} {Physical review letters}\
  }\textbf {\bibinfo {volume} {98}},\ \bibinfo {pages} {167203} (\bibinfo
  {year} {2007})}\BibitemShut {NoStop}%
\bibitem [{\citenamefont {Carmon}\ \emph {et~al.}(2005)\citenamefont {Carmon},
  \citenamefont {Rokhsari}, \citenamefont {Yang}, \citenamefont {Kippenberg},\
  and\ \citenamefont {Vahala}}]{carmon2005temporal}%
  \BibitemOpen
  \bibfield  {author} {\bibinfo {author} {\bibfnamefont {T.}~\bibnamefont
  {Carmon}}, \bibinfo {author} {\bibfnamefont {H.}~\bibnamefont {Rokhsari}},
  \bibinfo {author} {\bibfnamefont {L.}~\bibnamefont {Yang}}, \bibinfo {author}
  {\bibfnamefont {T.~J.}\ \bibnamefont {Kippenberg}},\ and\ \bibinfo {author}
  {\bibfnamefont {K.~J.}\ \bibnamefont {Vahala}},\ }\bibfield  {title}
  {\bibinfo {title} {Temporal behavior of radiation-pressure-induced vibrations
  of an optical microcavity phonon mode},\ }\href@noop {} {\bibfield  {journal}
  {\bibinfo  {journal} {Physical review letters}\ }\textbf {\bibinfo {volume}
  {94}},\ \bibinfo {pages} {223902} (\bibinfo {year} {2005})}\BibitemShut
  {NoStop}%
\bibitem [{\citenamefont {Metzger}\ \emph {et~al.}(2008)\citenamefont
  {Metzger}, \citenamefont {Ludwig}, \citenamefont {Neuenhahn}, \citenamefont
  {Ortlieb}, \citenamefont {Favero}, \citenamefont {Karrai},\ and\
  \citenamefont {Marquardt}}]{metzger2008self}%
  \BibitemOpen
  \bibfield  {author} {\bibinfo {author} {\bibfnamefont {C.}~\bibnamefont
  {Metzger}}, \bibinfo {author} {\bibfnamefont {M.}~\bibnamefont {Ludwig}},
  \bibinfo {author} {\bibfnamefont {C.}~\bibnamefont {Neuenhahn}}, \bibinfo
  {author} {\bibfnamefont {A.}~\bibnamefont {Ortlieb}}, \bibinfo {author}
  {\bibfnamefont {I.}~\bibnamefont {Favero}}, \bibinfo {author} {\bibfnamefont
  {.~f.~K.}\ \bibnamefont {Karrai}},\ and\ \bibinfo {author} {\bibfnamefont
  {F.}~\bibnamefont {Marquardt}},\ }\bibfield  {title} {\bibinfo {title}
  {Self-induced oscillations in an optomechanical system driven by bolometric
  backaction},\ }\href@noop {} {\bibfield  {journal} {\bibinfo  {journal}
  {Physical review letters}\ }\textbf {\bibinfo {volume} {101}},\ \bibinfo
  {pages} {133903} (\bibinfo {year} {2008})}\BibitemShut {NoStop}%
\bibitem [{\citenamefont {Marino}\ and\ \citenamefont
  {Marin}(2011)}]{marino2011chaotically}%
  \BibitemOpen
  \bibfield  {author} {\bibinfo {author} {\bibfnamefont {F.}~\bibnamefont
  {Marino}}\ and\ \bibinfo {author} {\bibfnamefont {F.}~\bibnamefont {Marin}},\
  }\bibfield  {title} {\bibinfo {title} {Chaotically spiking attractors in
  suspended-mirror optical cavities},\ }\href@noop {} {\bibfield  {journal}
  {\bibinfo  {journal} {Physical Review E—Statistical, Nonlinear, and Soft
  Matter Physics}\ }\textbf {\bibinfo {volume} {83}},\ \bibinfo {pages}
  {015202} (\bibinfo {year} {2011})}\BibitemShut {NoStop}%
\bibitem [{\citenamefont {Hollander}\ and\ \citenamefont
  {Gottlieb}(2012)}]{hollander2012self}%
  \BibitemOpen
  \bibfield  {author} {\bibinfo {author} {\bibfnamefont {E.}~\bibnamefont
  {Hollander}}\ and\ \bibinfo {author} {\bibfnamefont {O.}~\bibnamefont
  {Gottlieb}},\ }\bibfield  {title} {\bibinfo {title} {Self-excited chaotic
  dynamics of a nonlinear thermo-visco-elastic system that is subject to laser
  irradiation},\ }\href@noop {} {\bibfield  {journal} {\bibinfo  {journal}
  {Applied Physics Letters}\ }\textbf {\bibinfo {volume} {101}} (\bibinfo
  {year} {2012})}\BibitemShut {NoStop}%
\bibitem [{\citenamefont {Marino}\ and\ \citenamefont
  {Marin}(2013)}]{marino2013coexisting}%
  \BibitemOpen
  \bibfield  {author} {\bibinfo {author} {\bibfnamefont {F.}~\bibnamefont
  {Marino}}\ and\ \bibinfo {author} {\bibfnamefont {F.}~\bibnamefont {Marin}},\
  }\bibfield  {title} {\bibinfo {title} {Coexisting attractors and chaotic
  canard explosions in a slow-fast optomechanical system},\ }\href@noop {}
  {\bibfield  {journal} {\bibinfo  {journal} {Physical Review E—Statistical,
  Nonlinear, and Soft Matter Physics}\ }\textbf {\bibinfo {volume} {87}},\
  \bibinfo {pages} {052906} (\bibinfo {year} {2013})}\BibitemShut {NoStop}%
\bibitem [{\citenamefont {Wu}\ \emph {et~al.}(2017)\citenamefont {Wu},
  \citenamefont {Huang}, \citenamefont {Huang}, \citenamefont {Zhou},
  \citenamefont {Yang}, \citenamefont {Liu}, \citenamefont {Yu}, \citenamefont
  {Lo}, \citenamefont {Kwong}, \citenamefont {Duan} \emph
  {et~al.}}]{wu2017mesoscopic}%
  \BibitemOpen
  \bibfield  {author} {\bibinfo {author} {\bibfnamefont {J.}~\bibnamefont
  {Wu}}, \bibinfo {author} {\bibfnamefont {S.-W.}\ \bibnamefont {Huang}},
  \bibinfo {author} {\bibfnamefont {Y.}~\bibnamefont {Huang}}, \bibinfo
  {author} {\bibfnamefont {H.}~\bibnamefont {Zhou}}, \bibinfo {author}
  {\bibfnamefont {J.}~\bibnamefont {Yang}}, \bibinfo {author} {\bibfnamefont
  {J.-M.}\ \bibnamefont {Liu}}, \bibinfo {author} {\bibfnamefont
  {M.}~\bibnamefont {Yu}}, \bibinfo {author} {\bibfnamefont {G.}~\bibnamefont
  {Lo}}, \bibinfo {author} {\bibfnamefont {D.-L.}\ \bibnamefont {Kwong}},
  \bibinfo {author} {\bibfnamefont {S.}~\bibnamefont {Duan}}, \emph {et~al.},\
  }\bibfield  {title} {\bibinfo {title} {Mesoscopic chaos mediated by drude
  electron-hole plasma in silicon optomechanical oscillators},\ }\href@noop {}
  {\bibfield  {journal} {\bibinfo  {journal} {Nature communications}\ }\textbf
  {\bibinfo {volume} {8}},\ \bibinfo {pages} {15570} (\bibinfo {year}
  {2017})}\BibitemShut {NoStop}%
\bibitem [{\citenamefont {Navarro-Urrios}\ \emph {et~al.}(2017)\citenamefont
  {Navarro-Urrios}, \citenamefont {Capuj}, \citenamefont {Colombano},
  \citenamefont {Garc{\'\i}a}, \citenamefont {Sledzinska}, \citenamefont
  {Alzina}, \citenamefont {Griol}, \citenamefont {Mart{\'\i}nez},\ and\
  \citenamefont {Sotomayor-Torres}}]{navarro2017nonlinear}%
  \BibitemOpen
  \bibfield  {author} {\bibinfo {author} {\bibfnamefont {D.}~\bibnamefont
  {Navarro-Urrios}}, \bibinfo {author} {\bibfnamefont {N.~E.}\ \bibnamefont
  {Capuj}}, \bibinfo {author} {\bibfnamefont {M.~F.}\ \bibnamefont
  {Colombano}}, \bibinfo {author} {\bibfnamefont {P.~D.}\ \bibnamefont
  {Garc{\'\i}a}}, \bibinfo {author} {\bibfnamefont {M.}~\bibnamefont
  {Sledzinska}}, \bibinfo {author} {\bibfnamefont {F.}~\bibnamefont {Alzina}},
  \bibinfo {author} {\bibfnamefont {A.}~\bibnamefont {Griol}}, \bibinfo
  {author} {\bibfnamefont {A.}~\bibnamefont {Mart{\'\i}nez}},\ and\ \bibinfo
  {author} {\bibfnamefont {C.~M.}\ \bibnamefont {Sotomayor-Torres}},\
  }\bibfield  {title} {\bibinfo {title} {Nonlinear dynamics and chaos in an
  optomechanical beam},\ }\href@noop {} {\bibfield  {journal} {\bibinfo
  {journal} {Nature communications}\ }\textbf {\bibinfo {volume} {8}},\
  \bibinfo {pages} {14965} (\bibinfo {year} {2017})}\BibitemShut {NoStop}%
\bibitem [{\citenamefont {Gil-Santos}\ \emph {et~al.}(2017)\citenamefont
  {Gil-Santos}, \citenamefont {Labousse}, \citenamefont {Baker}, \citenamefont
  {Goetschy}, \citenamefont {Hease}, \citenamefont {Gomez}, \citenamefont
  {Lema{\^\i}tre}, \citenamefont {Leo}, \citenamefont {Ciuti},\ and\
  \citenamefont {Favero}}]{gil2017light}%
  \BibitemOpen
  \bibfield  {author} {\bibinfo {author} {\bibfnamefont {E.}~\bibnamefont
  {Gil-Santos}}, \bibinfo {author} {\bibfnamefont {M.}~\bibnamefont
  {Labousse}}, \bibinfo {author} {\bibfnamefont {C.}~\bibnamefont {Baker}},
  \bibinfo {author} {\bibfnamefont {A.}~\bibnamefont {Goetschy}}, \bibinfo
  {author} {\bibfnamefont {W.}~\bibnamefont {Hease}}, \bibinfo {author}
  {\bibfnamefont {C.}~\bibnamefont {Gomez}}, \bibinfo {author} {\bibfnamefont
  {A.}~\bibnamefont {Lema{\^\i}tre}}, \bibinfo {author} {\bibfnamefont
  {G.}~\bibnamefont {Leo}}, \bibinfo {author} {\bibfnamefont {C.}~\bibnamefont
  {Ciuti}},\ and\ \bibinfo {author} {\bibfnamefont {I.}~\bibnamefont
  {Favero}},\ }\bibfield  {title} {\bibinfo {title} {Light-mediated cascaded
  locking of multiple nano-optomechanical oscillators},\ }\href@noop {}
  {\bibfield  {journal} {\bibinfo  {journal} {Physical review letters}\
  }\textbf {\bibinfo {volume} {118}},\ \bibinfo {pages} {063605} (\bibinfo
  {year} {2017})}\BibitemShut {NoStop}%
\bibitem [{\citenamefont {Sciamanna}(2016)}]{sciamanna2016vibrations}%
  \BibitemOpen
  \bibfield  {author} {\bibinfo {author} {\bibfnamefont {M.}~\bibnamefont
  {Sciamanna}},\ }\bibfield  {title} {\bibinfo {title} {Vibrations copying
  optical chaos},\ }\href@noop {} {\bibfield  {journal} {\bibinfo  {journal}
  {Nature Photonics}\ }\textbf {\bibinfo {volume} {10}},\ \bibinfo {pages}
  {366} (\bibinfo {year} {2016})}\BibitemShut {NoStop}%
\bibitem [{\citenamefont {Ren}\ \emph {et~al.}(2022)\citenamefont {Ren},
  \citenamefont {Shah}, \citenamefont {Pfeifer}, \citenamefont {Brendel},
  \citenamefont {Peano}, \citenamefont {Marquardt},\ and\ \citenamefont
  {Painter}}]{ren2022topological}%
  \BibitemOpen
  \bibfield  {author} {\bibinfo {author} {\bibfnamefont {H.}~\bibnamefont
  {Ren}}, \bibinfo {author} {\bibfnamefont {T.}~\bibnamefont {Shah}}, \bibinfo
  {author} {\bibfnamefont {H.}~\bibnamefont {Pfeifer}}, \bibinfo {author}
  {\bibfnamefont {C.}~\bibnamefont {Brendel}}, \bibinfo {author} {\bibfnamefont
  {V.}~\bibnamefont {Peano}}, \bibinfo {author} {\bibfnamefont
  {F.}~\bibnamefont {Marquardt}},\ and\ \bibinfo {author} {\bibfnamefont
  {O.}~\bibnamefont {Painter}},\ }\bibfield  {title} {\bibinfo {title}
  {Topological phonon transport in an optomechanical system},\ }\href@noop {}
  {\bibfield  {journal} {\bibinfo  {journal} {Nature communications}\ }\textbf
  {\bibinfo {volume} {13}},\ \bibinfo {pages} {3476} (\bibinfo {year}
  {2022})}\BibitemShut {NoStop}%
\bibitem [{\citenamefont {Mondal}\ \emph {et~al.}(2024)\citenamefont {Mondal},
  \citenamefont {Baptista},\ and\ \citenamefont {Debnath}}]{mondal2024chaotic}%
  \BibitemOpen
  \bibfield  {author} {\bibinfo {author} {\bibfnamefont {S.}~\bibnamefont
  {Mondal}}, \bibinfo {author} {\bibfnamefont {M.~S.}\ \bibnamefont
  {Baptista}},\ and\ \bibinfo {author} {\bibfnamefont {K.}~\bibnamefont
  {Debnath}},\ }\bibfield  {title} {\bibinfo {title} {Chaotic dynamics under
  the influence of a synthetic magnetic field in an optomechanical system},\
  }\href@noop {} {\bibfield  {journal} {\bibinfo  {journal} {Physical Review
  A}\ }\textbf {\bibinfo {volume} {110}},\ \bibinfo {pages} {023509} (\bibinfo
  {year} {2024})}\BibitemShut {NoStop}%
\bibitem [{\citenamefont {Zhang}\ \emph {et~al.}(2020)\citenamefont {Zhang},
  \citenamefont {You},\ and\ \citenamefont {L{\"u}}}]{zhang2020intermittent}%
  \BibitemOpen
  \bibfield  {author} {\bibinfo {author} {\bibfnamefont {D.-W.}\ \bibnamefont
  {Zhang}}, \bibinfo {author} {\bibfnamefont {C.}~\bibnamefont {You}},\ and\
  \bibinfo {author} {\bibfnamefont {X.-Y.}\ \bibnamefont {L{\"u}}},\ }\bibfield
   {title} {\bibinfo {title} {Intermittent chaos in cavity optomechanics},\
  }\href@noop {} {\bibfield  {journal} {\bibinfo  {journal} {Physical Review
  A}\ }\textbf {\bibinfo {volume} {101}},\ \bibinfo {pages} {053851} (\bibinfo
  {year} {2020})}\BibitemShut {NoStop}%
\bibitem [{\citenamefont {L{\"u}}\ \emph {et~al.}(2015)\citenamefont {L{\"u}},
  \citenamefont {Jing}, \citenamefont {Ma},\ and\ \citenamefont
  {Wu}}]{lu2015pt}%
  \BibitemOpen
  \bibfield  {author} {\bibinfo {author} {\bibfnamefont {X.-Y.}\ \bibnamefont
  {L{\"u}}}, \bibinfo {author} {\bibfnamefont {H.}~\bibnamefont {Jing}},
  \bibinfo {author} {\bibfnamefont {J.-Y.}\ \bibnamefont {Ma}},\ and\ \bibinfo
  {author} {\bibfnamefont {Y.}~\bibnamefont {Wu}},\ }\bibfield  {title}
  {\bibinfo {title} {Pt-symmetry-breaking chaos in optomechanics},\ }\href@noop
  {} {\bibfield  {journal} {\bibinfo  {journal} {Physical review letters}\
  }\textbf {\bibinfo {volume} {114}},\ \bibinfo {pages} {253601} (\bibinfo
  {year} {2015})}\BibitemShut {NoStop}%
\bibitem [{\citenamefont {Ma}\ \emph {et~al.}(2014)\citenamefont {Ma},
  \citenamefont {You}, \citenamefont {Si}, \citenamefont {Xiong}, \citenamefont
  {Li}, \citenamefont {Yang},\ and\ \citenamefont {Wu}}]{ma2014formation}%
  \BibitemOpen
  \bibfield  {author} {\bibinfo {author} {\bibfnamefont {J.}~\bibnamefont
  {Ma}}, \bibinfo {author} {\bibfnamefont {C.}~\bibnamefont {You}}, \bibinfo
  {author} {\bibfnamefont {L.-G.}\ \bibnamefont {Si}}, \bibinfo {author}
  {\bibfnamefont {H.}~\bibnamefont {Xiong}}, \bibinfo {author} {\bibfnamefont
  {J.}~\bibnamefont {Li}}, \bibinfo {author} {\bibfnamefont {X.}~\bibnamefont
  {Yang}},\ and\ \bibinfo {author} {\bibfnamefont {Y.}~\bibnamefont {Wu}},\
  }\bibfield  {title} {\bibinfo {title} {Formation and manipulation of
  optomechanical chaos via a bichromatic driving},\ }\href@noop {} {\bibfield
  {journal} {\bibinfo  {journal} {Physical Review A}\ }\textbf {\bibinfo
  {volume} {90}},\ \bibinfo {pages} {043839} (\bibinfo {year}
  {2014})}\BibitemShut {NoStop}%
\bibitem [{\citenamefont {Yang}\ \emph {et~al.}(2019)\citenamefont {Yang},
  \citenamefont {Miranowicz}, \citenamefont {Liu}, \citenamefont {Xia},\ and\
  \citenamefont {Nori}}]{yang2019chaotic}%
  \BibitemOpen
  \bibfield  {author} {\bibinfo {author} {\bibfnamefont {N.}~\bibnamefont
  {Yang}}, \bibinfo {author} {\bibfnamefont {A.}~\bibnamefont {Miranowicz}},
  \bibinfo {author} {\bibfnamefont {Y.-C.}\ \bibnamefont {Liu}}, \bibinfo
  {author} {\bibfnamefont {K.}~\bibnamefont {Xia}},\ and\ \bibinfo {author}
  {\bibfnamefont {F.}~\bibnamefont {Nori}},\ }\bibfield  {title} {\bibinfo
  {title} {Chaotic synchronization of two optical cavity modes in
  optomechanical systems},\ }\href@noop {} {\bibfield  {journal} {\bibinfo
  {journal} {Scientific Reports}\ }\textbf {\bibinfo {volume} {9}},\ \bibinfo
  {pages} {15874} (\bibinfo {year} {2019})}\BibitemShut {NoStop}%
\bibitem [{\citenamefont {Monifi}\ \emph {et~al.}(2016)\citenamefont {Monifi},
  \citenamefont {Zhang}, \citenamefont {{\"O}zdemir}, \citenamefont {Peng},
  \citenamefont {Liu}, \citenamefont {Bo}, \citenamefont {Nori},\ and\
  \citenamefont {Yang}}]{monifi2016optomechanically}%
  \BibitemOpen
  \bibfield  {author} {\bibinfo {author} {\bibfnamefont {F.}~\bibnamefont
  {Monifi}}, \bibinfo {author} {\bibfnamefont {J.}~\bibnamefont {Zhang}},
  \bibinfo {author} {\bibfnamefont {{\c{S}}.~K.}\ \bibnamefont {{\"O}zdemir}},
  \bibinfo {author} {\bibfnamefont {B.}~\bibnamefont {Peng}}, \bibinfo {author}
  {\bibfnamefont {Y.-x.}\ \bibnamefont {Liu}}, \bibinfo {author} {\bibfnamefont
  {F.}~\bibnamefont {Bo}}, \bibinfo {author} {\bibfnamefont {F.}~\bibnamefont
  {Nori}},\ and\ \bibinfo {author} {\bibfnamefont {L.}~\bibnamefont {Yang}},\
  }\bibfield  {title} {\bibinfo {title} {Optomechanically induced stochastic
  resonance and chaos transfer between optical fields},\ }\href@noop {}
  {\bibfield  {journal} {\bibinfo  {journal} {Nature Photonics}\ }\textbf
  {\bibinfo {volume} {10}},\ \bibinfo {pages} {399} (\bibinfo {year}
  {2016})}\BibitemShut {NoStop}%
\bibitem [{\citenamefont {Madiot}\ \emph {et~al.}(2021)\citenamefont {Madiot},
  \citenamefont {Correia}, \citenamefont {Barbay},\ and\ \citenamefont
  {Braive}}]{madiot2021bichromatic}%
  \BibitemOpen
  \bibfield  {author} {\bibinfo {author} {\bibfnamefont {G.}~\bibnamefont
  {Madiot}}, \bibinfo {author} {\bibfnamefont {F.}~\bibnamefont {Correia}},
  \bibinfo {author} {\bibfnamefont {S.}~\bibnamefont {Barbay}},\ and\ \bibinfo
  {author} {\bibfnamefont {R.}~\bibnamefont {Braive}},\ }\bibfield  {title}
  {\bibinfo {title} {Bichromatic synchronized chaos in driven coupled
  electro-optomechanical nanoresonators},\ }\href@noop {} {\bibfield  {journal}
  {\bibinfo  {journal} {Physical Review A}\ }\textbf {\bibinfo {volume}
  {104}},\ \bibinfo {pages} {023525} (\bibinfo {year} {2021})}\BibitemShut
  {NoStop}%
\bibitem [{\citenamefont {Pikovsky}\ \emph {et~al.}(2001)\citenamefont
  {Pikovsky}, \citenamefont {Rosenblum},\ and\ \citenamefont
  {Kurths}}]{pikovsky2001synchronization}%
  \BibitemOpen
  \bibfield  {author} {\bibinfo {author} {\bibfnamefont {A.}~\bibnamefont
  {Pikovsky}}, \bibinfo {author} {\bibfnamefont {M.}~\bibnamefont
  {Rosenblum}},\ and\ \bibinfo {author} {\bibfnamefont {J.}~\bibnamefont
  {Kurths}},\ }\bibfield  {title} {\bibinfo {title} {Synchronization},\
  }\href@noop {} {\bibfield  {journal} {\bibinfo  {journal} {Cambridge
  university press}\ }\textbf {\bibinfo {volume} {12}} (\bibinfo {year}
  {2001})}\BibitemShut {NoStop}%
\bibitem [{\citenamefont {Boccaletti}\ \emph {et~al.}(2002)\citenamefont
  {Boccaletti}, \citenamefont {Kurths}, \citenamefont {Osipov}, \citenamefont
  {Valladares},\ and\ \citenamefont {Zhou}}]{boccaletti2002synchronization}%
  \BibitemOpen
  \bibfield  {author} {\bibinfo {author} {\bibfnamefont {S.}~\bibnamefont
  {Boccaletti}}, \bibinfo {author} {\bibfnamefont {J.}~\bibnamefont {Kurths}},
  \bibinfo {author} {\bibfnamefont {G.}~\bibnamefont {Osipov}}, \bibinfo
  {author} {\bibfnamefont {D.}~\bibnamefont {Valladares}},\ and\ \bibinfo
  {author} {\bibfnamefont {C.}~\bibnamefont {Zhou}},\ }\bibfield  {title}
  {\bibinfo {title} {The synchronization of chaotic systems},\ }\href@noop {}
  {\bibfield  {journal} {\bibinfo  {journal} {Physics reports}\ }\textbf
  {\bibinfo {volume} {366}},\ \bibinfo {pages} {1} (\bibinfo {year}
  {2002})}\BibitemShut {NoStop}%
\bibitem [{\citenamefont {Deniz~Eroglu}\ and\ \citenamefont
  {Pereira}(2017)}]{eroglu2017}%
  \BibitemOpen
  \bibfield  {author} {\bibinfo {author} {\bibfnamefont {J.~S. W.~L.}\
  \bibnamefont {Deniz~Eroglu}}\ and\ \bibinfo {author} {\bibfnamefont
  {T.}~\bibnamefont {Pereira}},\ }\bibfield  {title} {\bibinfo {title}
  {Synchronisation of chaos and its applications},\ }\href@noop {} {\bibfield
  {journal} {\bibinfo  {journal} {Contemporary Physics}\ }\textbf {\bibinfo
  {volume} {58}},\ \bibinfo {pages} {207} (\bibinfo {year} {2017})}\BibitemShut
  {NoStop}%
\bibitem [{\citenamefont {Locquet}\ \emph {et~al.}(2002)\citenamefont
  {Locquet}, \citenamefont {Masoller},\ and\ \citenamefont
  {Mirasso}}]{locquet2002synchronization}%
  \BibitemOpen
  \bibfield  {author} {\bibinfo {author} {\bibfnamefont {A.}~\bibnamefont
  {Locquet}}, \bibinfo {author} {\bibfnamefont {C.}~\bibnamefont {Masoller}},\
  and\ \bibinfo {author} {\bibfnamefont {C.~R.}\ \bibnamefont {Mirasso}},\
  }\bibfield  {title} {\bibinfo {title} {Synchronization regimes of
  optical-feedback-induced chaos in unidirectionally coupled semiconductor
  lasers},\ }\href@noop {} {\bibfield  {journal} {\bibinfo  {journal} {Physical
  Review E}\ }\textbf {\bibinfo {volume} {65}},\ \bibinfo {pages} {056205}
  (\bibinfo {year} {2002})}\BibitemShut {NoStop}%
\bibitem [{\citenamefont {Murakami}(2002)}]{murakami2002synchronization}%
  \BibitemOpen
  \bibfield  {author} {\bibinfo {author} {\bibfnamefont {A.}~\bibnamefont
  {Murakami}},\ }\bibfield  {title} {\bibinfo {title} {Synchronization of chaos
  due to linear response in optically driven semiconductor lasers},\
  }\href@noop {} {\bibfield  {journal} {\bibinfo  {journal} {Physical Review
  E}\ }\textbf {\bibinfo {volume} {65}},\ \bibinfo {pages} {056617} (\bibinfo
  {year} {2002})}\BibitemShut {NoStop}%
\bibitem [{\citenamefont {Eichler}\ \emph {et~al.}(2018)\citenamefont
  {Eichler}, \citenamefont {Eichler}, \citenamefont {Lux}, \citenamefont
  {Eichler}, \citenamefont {Eichler},\ and\ \citenamefont
  {Lux}}]{eichler2018semiconductor}%
  \BibitemOpen
  \bibfield  {author} {\bibinfo {author} {\bibfnamefont {H.~J.}\ \bibnamefont
  {Eichler}}, \bibinfo {author} {\bibfnamefont {J.}~\bibnamefont {Eichler}},
  \bibinfo {author} {\bibfnamefont {O.}~\bibnamefont {Lux}}, \bibinfo {author}
  {\bibfnamefont {H.~J.}\ \bibnamefont {Eichler}}, \bibinfo {author}
  {\bibfnamefont {J.}~\bibnamefont {Eichler}},\ and\ \bibinfo {author}
  {\bibfnamefont {O.}~\bibnamefont {Lux}},\ }\bibfield  {title} {\bibinfo
  {title} {Semiconductor lasers},\ }\href@noop {} {\bibfield  {journal}
  {\bibinfo  {journal} {Lasers: Basics, Advances and Applications}\ ,\ \bibinfo
  {pages} {165}} (\bibinfo {year} {2018})}\BibitemShut {NoStop}%
\bibitem [{\citenamefont {Acharyya}\ and\ \citenamefont
  {Amritkar}(2012)}]{acharyya2012synchronization}%
  \BibitemOpen
  \bibfield  {author} {\bibinfo {author} {\bibfnamefont {S.}~\bibnamefont
  {Acharyya}}\ and\ \bibinfo {author} {\bibfnamefont {R.}~\bibnamefont
  {Amritkar}},\ }\bibfield  {title} {\bibinfo {title} {Synchronization of
  coupled nonidentical dynamical systems},\ }\href@noop {} {\bibfield
  {journal} {\bibinfo  {journal} {Europhysics Letters}\ }\textbf {\bibinfo
  {volume} {99}},\ \bibinfo {pages} {40005} (\bibinfo {year}
  {2012})}\BibitemShut {NoStop}%
\bibitem [{\citenamefont {Rulkov}\ \emph {et~al.}(1995)\citenamefont {Rulkov},
  \citenamefont {Sushchik}, \citenamefont {Tsimring},\ and\ \citenamefont
  {Abarbanel}}]{rulkov1995generalized}%
  \BibitemOpen
  \bibfield  {author} {\bibinfo {author} {\bibfnamefont {N.~F.}\ \bibnamefont
  {Rulkov}}, \bibinfo {author} {\bibfnamefont {M.~M.}\ \bibnamefont
  {Sushchik}}, \bibinfo {author} {\bibfnamefont {L.~S.}\ \bibnamefont
  {Tsimring}},\ and\ \bibinfo {author} {\bibfnamefont {H.~D.}\ \bibnamefont
  {Abarbanel}},\ }\bibfield  {title} {\bibinfo {title} {Generalized
  synchronization of chaos in directionally coupled chaotic systems},\
  }\href@noop {} {\bibfield  {journal} {\bibinfo  {journal} {Physical Review
  E}\ }\textbf {\bibinfo {volume} {51}},\ \bibinfo {pages} {980} (\bibinfo
  {year} {1995})}\BibitemShut {NoStop}%
\bibitem [{\citenamefont {Dahms}\ \emph {et~al.}(2012)\citenamefont {Dahms},
  \citenamefont {Lehnert},\ and\ \citenamefont
  {Sch{\"o}ll}}]{dahms2012cluster}%
  \BibitemOpen
  \bibfield  {author} {\bibinfo {author} {\bibfnamefont {T.}~\bibnamefont
  {Dahms}}, \bibinfo {author} {\bibfnamefont {J.}~\bibnamefont {Lehnert}},\
  and\ \bibinfo {author} {\bibfnamefont {E.}~\bibnamefont {Sch{\"o}ll}},\
  }\bibfield  {title} {\bibinfo {title} {Cluster and group synchronization in
  delay-coupled networks},\ }\href@noop {} {\bibfield  {journal} {\bibinfo
  {journal} {Physical Review E—Statistical, Nonlinear, and Soft Matter
  Physics}\ }\textbf {\bibinfo {volume} {86}},\ \bibinfo {pages} {016202}
  (\bibinfo {year} {2012})}\BibitemShut {NoStop}%
\bibitem [{\citenamefont {Belykh}\ \emph {et~al.}(2003)\citenamefont {Belykh},
  \citenamefont {Belykh}, \citenamefont {Nevidin},\ and\ \citenamefont
  {Hasler}}]{belykh2003persistent}%
  \BibitemOpen
  \bibfield  {author} {\bibinfo {author} {\bibfnamefont {I.}~\bibnamefont
  {Belykh}}, \bibinfo {author} {\bibfnamefont {V.}~\bibnamefont {Belykh}},
  \bibinfo {author} {\bibfnamefont {K.}~\bibnamefont {Nevidin}},\ and\ \bibinfo
  {author} {\bibfnamefont {M.}~\bibnamefont {Hasler}},\ }\bibfield  {title}
  {\bibinfo {title} {Persistent clusters in lattices of coupled nonidentical
  chaotic systems},\ }\href@noop {} {\bibfield  {journal} {\bibinfo  {journal}
  {Chaos: An Interdisciplinary Journal of Nonlinear Science}\ }\textbf
  {\bibinfo {volume} {13}},\ \bibinfo {pages} {165} (\bibinfo {year}
  {2003})}\BibitemShut {NoStop}%
\bibitem [{\citenamefont {Pecora}\ \emph {et~al.}(2014)\citenamefont {Pecora},
  \citenamefont {Sorrentino}, \citenamefont {Hagerstrom}, \citenamefont
  {Murphy},\ and\ \citenamefont {Roy}}]{pecora2014cluster}%
  \BibitemOpen
  \bibfield  {author} {\bibinfo {author} {\bibfnamefont {L.~M.}\ \bibnamefont
  {Pecora}}, \bibinfo {author} {\bibfnamefont {F.}~\bibnamefont {Sorrentino}},
  \bibinfo {author} {\bibfnamefont {A.~M.}\ \bibnamefont {Hagerstrom}},
  \bibinfo {author} {\bibfnamefont {T.~E.}\ \bibnamefont {Murphy}},\ and\
  \bibinfo {author} {\bibfnamefont {R.}~\bibnamefont {Roy}},\ }\bibfield
  {title} {\bibinfo {title} {Cluster synchronization and isolated
  desynchronization in complex networks with symmetries},\ }\href@noop {}
  {\bibfield  {journal} {\bibinfo  {journal} {Nature communications}\ }\textbf
  {\bibinfo {volume} {5}},\ \bibinfo {pages} {4079} (\bibinfo {year}
  {2014})}\BibitemShut {NoStop}%
\bibitem [{\citenamefont {Zhou}\ and\ \citenamefont
  {Kurths}(2002)}]{zhou2002noise}%
  \BibitemOpen
  \bibfield  {author} {\bibinfo {author} {\bibfnamefont {C.}~\bibnamefont
  {Zhou}}\ and\ \bibinfo {author} {\bibfnamefont {J.}~\bibnamefont {Kurths}},\
  }\bibfield  {title} {\bibinfo {title} {Noise-induced phase synchronization
  and synchronization transitions in chaotic oscillators},\ }\href@noop {}
  {\bibfield  {journal} {\bibinfo  {journal} {Physical review letters}\
  }\textbf {\bibinfo {volume} {88}},\ \bibinfo {pages} {230602} (\bibinfo
  {year} {2002})}\BibitemShut {NoStop}%
\bibitem [{\citenamefont {Kiss}\ and\ \citenamefont
  {Hudson}(2002)}]{kiss2002phase}%
  \BibitemOpen
  \bibfield  {author} {\bibinfo {author} {\bibfnamefont {I.~Z.}\ \bibnamefont
  {Kiss}}\ and\ \bibinfo {author} {\bibfnamefont {J.~L.}\ \bibnamefont
  {Hudson}},\ }\bibfield  {title} {\bibinfo {title} {Phase synchronization of
  nonidentical chaotic electrochemical oscillators},\ }\href@noop {} {\bibfield
   {journal} {\bibinfo  {journal} {Physical Chemistry Chemical Physics}\
  }\textbf {\bibinfo {volume} {4}},\ \bibinfo {pages} {2638} (\bibinfo {year}
  {2002})}\BibitemShut {NoStop}%
\bibitem [{\citenamefont {Chen}(2005)}]{chen2005synchronization}%
  \BibitemOpen
  \bibfield  {author} {\bibinfo {author} {\bibfnamefont {H.-K.}\ \bibnamefont
  {Chen}},\ }\bibfield  {title} {\bibinfo {title} {Synchronization of two
  different chaotic systems: a new system and each of the dynamical systems
  lorenz, chen and l{\"u}},\ }\href@noop {} {\bibfield  {journal} {\bibinfo
  {journal} {Chaos, Solitons \& Fractals}\ }\textbf {\bibinfo {volume} {25}},\
  \bibinfo {pages} {1049} (\bibinfo {year} {2005})}\BibitemShut {NoStop}%
\bibitem [{\citenamefont {Yassen}(2005)}]{yassen2005chaos}%
  \BibitemOpen
  \bibfield  {author} {\bibinfo {author} {\bibfnamefont {M.}~\bibnamefont
  {Yassen}},\ }\bibfield  {title} {\bibinfo {title} {Chaos synchronization
  between two different chaotic systems using active control},\ }\href@noop {}
  {\bibfield  {journal} {\bibinfo  {journal} {Chaos, Solitons \& Fractals}\
  }\textbf {\bibinfo {volume} {23}},\ \bibinfo {pages} {131} (\bibinfo {year}
  {2005})}\BibitemShut {NoStop}%
\bibitem [{\citenamefont {Williams}\ \emph {et~al.}(2013)\citenamefont
  {Williams}, \citenamefont {Murphy}, \citenamefont {Roy}, \citenamefont
  {Sorrentino}, \citenamefont {Dahms},\ and\ \citenamefont
  {Sch{\"o}ll}}]{williams2013experimental}%
  \BibitemOpen
  \bibfield  {author} {\bibinfo {author} {\bibfnamefont {C.~R.}\ \bibnamefont
  {Williams}}, \bibinfo {author} {\bibfnamefont {T.~E.}\ \bibnamefont
  {Murphy}}, \bibinfo {author} {\bibfnamefont {R.}~\bibnamefont {Roy}},
  \bibinfo {author} {\bibfnamefont {F.}~\bibnamefont {Sorrentino}}, \bibinfo
  {author} {\bibfnamefont {T.}~\bibnamefont {Dahms}},\ and\ \bibinfo {author}
  {\bibfnamefont {E.}~\bibnamefont {Sch{\"o}ll}},\ }\bibfield  {title}
  {\bibinfo {title} {Experimental observations of group synchrony in a system
  of chaotic optoelectronic oscillators},\ }\href@noop {} {\bibfield  {journal}
  {\bibinfo  {journal} {Physical review letters}\ }\textbf {\bibinfo {volume}
  {110}},\ \bibinfo {pages} {064104} (\bibinfo {year} {2013})}\BibitemShut
  {NoStop}%
\bibitem [{\citenamefont {Landsman}\ and\ \citenamefont
  {Schwartz}(2007)}]{landsman2007complete}%
  \BibitemOpen
  \bibfield  {author} {\bibinfo {author} {\bibfnamefont {A.~S.}\ \bibnamefont
  {Landsman}}\ and\ \bibinfo {author} {\bibfnamefont {I.~B.}\ \bibnamefont
  {Schwartz}},\ }\bibfield  {title} {\bibinfo {title} {Complete chaotic
  synchronization in mutually coupled time-delay systems},\ }\href@noop {}
  {\bibfield  {journal} {\bibinfo  {journal} {Physical Review E—Statistical,
  Nonlinear, and Soft Matter Physics}\ }\textbf {\bibinfo {volume} {75}},\
  \bibinfo {pages} {026201} (\bibinfo {year} {2007})}\BibitemShut {NoStop}%
\bibitem [{\citenamefont {Zamora-Munt}\ \emph {et~al.}(2010)\citenamefont
  {Zamora-Munt}, \citenamefont {Masoller}, \citenamefont {Garcia-Ojalvo},\ and\
  \citenamefont {Roy}}]{zamora2010crowd}%
  \BibitemOpen
  \bibfield  {author} {\bibinfo {author} {\bibfnamefont {J.}~\bibnamefont
  {Zamora-Munt}}, \bibinfo {author} {\bibfnamefont {C.}~\bibnamefont
  {Masoller}}, \bibinfo {author} {\bibfnamefont {J.}~\bibnamefont
  {Garcia-Ojalvo}},\ and\ \bibinfo {author} {\bibfnamefont {R.}~\bibnamefont
  {Roy}},\ }\bibfield  {title} {\bibinfo {title} {Crowd synchrony and quorum
  sensing in delay-coupled lasers},\ }\href@noop {} {\bibfield  {journal}
  {\bibinfo  {journal} {Physical review letters}\ }\textbf {\bibinfo {volume}
  {105}},\ \bibinfo {pages} {264101} (\bibinfo {year} {2010})}\BibitemShut
  {NoStop}%
\bibitem [{\citenamefont {Kippenberg}\ \emph {et~al.}(2005)\citenamefont
  {Kippenberg}, \citenamefont {Rokhsari}, \citenamefont {Carmon}, \citenamefont
  {Scherer},\ and\ \citenamefont {Vahala}}]{kippenberg2005analysis}%
  \BibitemOpen
  \bibfield  {author} {\bibinfo {author} {\bibfnamefont {T.}~\bibnamefont
  {Kippenberg}}, \bibinfo {author} {\bibfnamefont {H.}~\bibnamefont
  {Rokhsari}}, \bibinfo {author} {\bibfnamefont {T.}~\bibnamefont {Carmon}},
  \bibinfo {author} {\bibfnamefont {A.}~\bibnamefont {Scherer}},\ and\ \bibinfo
  {author} {\bibfnamefont {K.}~\bibnamefont {Vahala}},\ }\bibfield  {title}
  {\bibinfo {title} {Analysis of radiation-pressure induced mechanical
  oscillation of an optical microcavity},\ }\href@noop {} {\bibfield  {journal}
  {\bibinfo  {journal} {Physical Review Letters}\ }\textbf {\bibinfo {volume}
  {95}},\ \bibinfo {pages} {033901} (\bibinfo {year} {2005})}\BibitemShut
  {NoStop}%
\bibitem [{\citenamefont {Schliesser}\ \emph {et~al.}(2008)\citenamefont
  {Schliesser}, \citenamefont {Anetsberger}, \citenamefont {Rivi{\`e}re},
  \citenamefont {Arcizet},\ and\ \citenamefont
  {Kippenberg}}]{schliesser2008high}%
  \BibitemOpen
  \bibfield  {author} {\bibinfo {author} {\bibfnamefont {A.}~\bibnamefont
  {Schliesser}}, \bibinfo {author} {\bibfnamefont {G.}~\bibnamefont
  {Anetsberger}}, \bibinfo {author} {\bibfnamefont {R.}~\bibnamefont
  {Rivi{\`e}re}}, \bibinfo {author} {\bibfnamefont {O.}~\bibnamefont
  {Arcizet}},\ and\ \bibinfo {author} {\bibfnamefont {T.~J.}\ \bibnamefont
  {Kippenberg}},\ }\bibfield  {title} {\bibinfo {title} {High-sensitivity
  monitoring of micromechanical vibration using optical whispering gallery mode
  resonators},\ }\href@noop {} {\bibfield  {journal} {\bibinfo  {journal} {New
  Journal of Physics}\ }\textbf {\bibinfo {volume} {10}},\ \bibinfo {pages}
  {095015} (\bibinfo {year} {2008})}\BibitemShut {NoStop}%
\bibitem [{\citenamefont {Weis}\ \emph {et~al.}(2010)\citenamefont {Weis},
  \citenamefont {Rivi{\`e}re}, \citenamefont {Del{\'e}glise}, \citenamefont
  {Gavartin}, \citenamefont {Arcizet}, \citenamefont {Schliesser},\ and\
  \citenamefont {Kippenberg}}]{weis2010optomechanically}%
  \BibitemOpen
  \bibfield  {author} {\bibinfo {author} {\bibfnamefont {S.}~\bibnamefont
  {Weis}}, \bibinfo {author} {\bibfnamefont {R.}~\bibnamefont {Rivi{\`e}re}},
  \bibinfo {author} {\bibfnamefont {S.}~\bibnamefont {Del{\'e}glise}}, \bibinfo
  {author} {\bibfnamefont {E.}~\bibnamefont {Gavartin}}, \bibinfo {author}
  {\bibfnamefont {O.}~\bibnamefont {Arcizet}}, \bibinfo {author} {\bibfnamefont
  {A.}~\bibnamefont {Schliesser}},\ and\ \bibinfo {author} {\bibfnamefont
  {T.~J.}\ \bibnamefont {Kippenberg}},\ }\bibfield  {title} {\bibinfo {title}
  {Optomechanically induced transparency},\ }\href@noop {} {\bibfield
  {journal} {\bibinfo  {journal} {Science}\ }\textbf {\bibinfo {volume}
  {330}},\ \bibinfo {pages} {1520} (\bibinfo {year} {2010})}\BibitemShut
  {NoStop}%
\bibitem [{\citenamefont {Walls}\ and\ \citenamefont
  {Milburn}(2008)}]{walls2008input}%
  \BibitemOpen
  \bibfield  {author} {\bibinfo {author} {\bibfnamefont {D.}~\bibnamefont
  {Walls}}\ and\ \bibinfo {author} {\bibfnamefont {G.~J.}\ \bibnamefont
  {Milburn}},\ }\bibfield  {title} {\bibinfo {title} {Input--output formulation
  of optical cavities},\ }in\ \href@noop {} {\emph {\bibinfo {booktitle}
  {Quantum optics}}}\ (\bibinfo  {publisher} {Springer},\ \bibinfo {year}
  {2008})\ pp.\ \bibinfo {pages} {127--141}\BibitemShut {NoStop}%
\bibitem [{\citenamefont {Ristic}\ \emph {et~al.}(2009)\citenamefont {Ristic},
  \citenamefont {Bhardwaj}, \citenamefont {Rodwell}, \citenamefont {Coldren},\
  and\ \citenamefont {Johansson}}]{ristic2009optical}%
  \BibitemOpen
  \bibfield  {author} {\bibinfo {author} {\bibfnamefont {S.}~\bibnamefont
  {Ristic}}, \bibinfo {author} {\bibfnamefont {A.}~\bibnamefont {Bhardwaj}},
  \bibinfo {author} {\bibfnamefont {M.~J.}\ \bibnamefont {Rodwell}}, \bibinfo
  {author} {\bibfnamefont {L.~A.}\ \bibnamefont {Coldren}},\ and\ \bibinfo
  {author} {\bibfnamefont {L.~A.}\ \bibnamefont {Johansson}},\ }\bibfield
  {title} {\bibinfo {title} {An optical phase-locked loop photonic integrated
  circuit},\ }\href@noop {} {\bibfield  {journal} {\bibinfo  {journal} {Journal
  of Lightwave Technology}\ }\textbf {\bibinfo {volume} {28}},\ \bibinfo
  {pages} {526} (\bibinfo {year} {2009})}\BibitemShut {NoStop}%
\bibitem [{\citenamefont {Arafin}\ \emph {et~al.}(2017)\citenamefont {Arafin},
  \citenamefont {Simsek}, \citenamefont {Lu}, \citenamefont {Rodwell},\ and\
  \citenamefont {Coldren}}]{arafin2017heterodyne}%
  \BibitemOpen
  \bibfield  {author} {\bibinfo {author} {\bibfnamefont {S.}~\bibnamefont
  {Arafin}}, \bibinfo {author} {\bibfnamefont {A.}~\bibnamefont {Simsek}},
  \bibinfo {author} {\bibfnamefont {M.}~\bibnamefont {Lu}}, \bibinfo {author}
  {\bibfnamefont {M.~J.}\ \bibnamefont {Rodwell}},\ and\ \bibinfo {author}
  {\bibfnamefont {L.~A.}\ \bibnamefont {Coldren}},\ }\bibfield  {title}
  {\bibinfo {title} {Heterodyne locking of a fully integrated optical
  phase-locked loop with on-chip modulators},\ }\href@noop {} {\bibfield
  {journal} {\bibinfo  {journal} {Optics Letters}\ }\textbf {\bibinfo {volume}
  {42}},\ \bibinfo {pages} {3745} (\bibinfo {year} {2017})}\BibitemShut
  {NoStop}%
\bibitem [{\citenamefont {Aspelmeyer}\ \emph {et~al.}(2014)\citenamefont
  {Aspelmeyer}, \citenamefont {Kippenberg},\ and\ \citenamefont
  {Marquardt}}]{aspelmeyer2014cavity}%
  \BibitemOpen
  \bibfield  {author} {\bibinfo {author} {\bibfnamefont {M.}~\bibnamefont
  {Aspelmeyer}}, \bibinfo {author} {\bibfnamefont {T.~J.}\ \bibnamefont
  {Kippenberg}},\ and\ \bibinfo {author} {\bibfnamefont {F.}~\bibnamefont
  {Marquardt}},\ }\bibfield  {title} {\bibinfo {title} {Cavity optomechanics},\
  }\href@noop {} {\bibfield  {journal} {\bibinfo  {journal} {Reviews of Modern
  Physics}\ }\textbf {\bibinfo {volume} {86}},\ \bibinfo {pages} {1391}
  (\bibinfo {year} {2014})}\BibitemShut {NoStop}%
\bibitem [{\citenamefont {Manolatou}\ \emph {et~al.}(1999)\citenamefont
  {Manolatou}, \citenamefont {Khan}, \citenamefont {Fan}, \citenamefont
  {Villeneuve}, \citenamefont {Haus},\ and\ \citenamefont
  {Joannopoulos}}]{784592}%
  \BibitemOpen
  \bibfield  {author} {\bibinfo {author} {\bibfnamefont {C.}~\bibnamefont
  {Manolatou}}, \bibinfo {author} {\bibfnamefont {M.}~\bibnamefont {Khan}},
  \bibinfo {author} {\bibfnamefont {S.}~\bibnamefont {Fan}}, \bibinfo {author}
  {\bibfnamefont {P.}~\bibnamefont {Villeneuve}}, \bibinfo {author}
  {\bibfnamefont {H.}~\bibnamefont {Haus}},\ and\ \bibinfo {author}
  {\bibfnamefont {J.}~\bibnamefont {Joannopoulos}},\ }\bibfield  {title}
  {\bibinfo {title} {Coupling of modes analysis of resonant channel add-drop
  filters},\ }\href@noop {} {\bibfield  {journal} {\bibinfo  {journal} {IEEE
  Journal of Quantum Electronics}\ }\textbf {\bibinfo {volume} {35}},\ \bibinfo
  {pages} {1322} (\bibinfo {year} {1999})}\BibitemShut {NoStop}%
\bibitem [{\citenamefont {McKenna}\ \emph {et~al.}(2019)\citenamefont
  {McKenna}, \citenamefont {Patel}, \citenamefont {Witmer}, \citenamefont
  {Van~Laer}, \citenamefont {Valery},\ and\ \citenamefont
  {Safavi-Naeini}}]{mckenna2019cryogenic}%
  \BibitemOpen
  \bibfield  {author} {\bibinfo {author} {\bibfnamefont {T.~P.}\ \bibnamefont
  {McKenna}}, \bibinfo {author} {\bibfnamefont {R.~N.}\ \bibnamefont {Patel}},
  \bibinfo {author} {\bibfnamefont {J.~D.}\ \bibnamefont {Witmer}}, \bibinfo
  {author} {\bibfnamefont {R.}~\bibnamefont {Van~Laer}}, \bibinfo {author}
  {\bibfnamefont {J.~A.}\ \bibnamefont {Valery}},\ and\ \bibinfo {author}
  {\bibfnamefont {A.~H.}\ \bibnamefont {Safavi-Naeini}},\ }\bibfield  {title}
  {\bibinfo {title} {Cryogenic packaging of an optomechanical crystal},\
  }\href@noop {} {\bibfield  {journal} {\bibinfo  {journal} {Optics express}\
  }\textbf {\bibinfo {volume} {27}},\ \bibinfo {pages} {28782} (\bibinfo {year}
  {2019})}\BibitemShut {NoStop}%
\bibitem [{\citenamefont {Kuhn}\ \emph {et~al.}(2014)\citenamefont {Kuhn},
  \citenamefont {Teissier}, \citenamefont {Neuhaus}, \citenamefont {Zerkani},
  \citenamefont {Van~Brackel}, \citenamefont {Del{\'e}glise}, \citenamefont
  {Briant}, \citenamefont {Cohadon}, \citenamefont {Heidmann}, \citenamefont
  {Michel} \emph {et~al.}}]{kuhn2014free}%
  \BibitemOpen
  \bibfield  {author} {\bibinfo {author} {\bibfnamefont {A.}~\bibnamefont
  {Kuhn}}, \bibinfo {author} {\bibfnamefont {J.}~\bibnamefont {Teissier}},
  \bibinfo {author} {\bibfnamefont {L.}~\bibnamefont {Neuhaus}}, \bibinfo
  {author} {\bibfnamefont {S.}~\bibnamefont {Zerkani}}, \bibinfo {author}
  {\bibfnamefont {E.}~\bibnamefont {Van~Brackel}}, \bibinfo {author}
  {\bibfnamefont {S.}~\bibnamefont {Del{\'e}glise}}, \bibinfo {author}
  {\bibfnamefont {T.}~\bibnamefont {Briant}}, \bibinfo {author} {\bibfnamefont
  {P.-F.}\ \bibnamefont {Cohadon}}, \bibinfo {author} {\bibfnamefont
  {A.}~\bibnamefont {Heidmann}}, \bibinfo {author} {\bibfnamefont
  {C.}~\bibnamefont {Michel}}, \emph {et~al.},\ }\bibfield  {title} {\bibinfo
  {title} {Free-space cavity optomechanics in a cryogenic environment},\
  }\href@noop {} {\bibfield  {journal} {\bibinfo  {journal} {Applied Physics
  Letters}\ }\textbf {\bibinfo {volume} {104}} (\bibinfo {year}
  {2014})}\BibitemShut {NoStop}%
\bibitem [{\citenamefont {Fu}\ \emph {et~al.}(2017)\citenamefont {Fu},
  \citenamefont {Shi}, \citenamefont {Feng}, \citenamefont {Zhang},
  \citenamefont {Yang}, \citenamefont {Xu}, \citenamefont {Zhu}, \citenamefont
  {Norwood},\ and\ \citenamefont {Peyghambarian}}]{fu2017review}%
  \BibitemOpen
  \bibfield  {author} {\bibinfo {author} {\bibfnamefont {S.}~\bibnamefont
  {Fu}}, \bibinfo {author} {\bibfnamefont {W.}~\bibnamefont {Shi}}, \bibinfo
  {author} {\bibfnamefont {Y.}~\bibnamefont {Feng}}, \bibinfo {author}
  {\bibfnamefont {L.}~\bibnamefont {Zhang}}, \bibinfo {author} {\bibfnamefont
  {Z.}~\bibnamefont {Yang}}, \bibinfo {author} {\bibfnamefont {S.}~\bibnamefont
  {Xu}}, \bibinfo {author} {\bibfnamefont {X.}~\bibnamefont {Zhu}}, \bibinfo
  {author} {\bibfnamefont {R.~A.}\ \bibnamefont {Norwood}},\ and\ \bibinfo
  {author} {\bibfnamefont {N.}~\bibnamefont {Peyghambarian}},\ }\bibfield
  {title} {\bibinfo {title} {Review of recent progress on single-frequency
  fiber lasers},\ }\href@noop {} {\bibfield  {journal} {\bibinfo  {journal}
  {Journal of the Optical Society of America B}\ }\textbf {\bibinfo {volume}
  {34}},\ \bibinfo {pages} {A49} (\bibinfo {year} {2017})}\BibitemShut
  {NoStop}%
\bibitem [{\citenamefont {Yang}\ \emph {et~al.}(2020)\citenamefont {Yang},
  \citenamefont {Yi}, \citenamefont {Ke}, \citenamefont {Zhuge}, \citenamefont
  {Yang},\ and\ \citenamefont {Hu}}]{yang2020chaotic}%
  \BibitemOpen
  \bibfield  {author} {\bibinfo {author} {\bibfnamefont {Z.}~\bibnamefont
  {Yang}}, \bibinfo {author} {\bibfnamefont {L.}~\bibnamefont {Yi}}, \bibinfo
  {author} {\bibfnamefont {J.}~\bibnamefont {Ke}}, \bibinfo {author}
  {\bibfnamefont {Q.}~\bibnamefont {Zhuge}}, \bibinfo {author} {\bibfnamefont
  {Y.}~\bibnamefont {Yang}},\ and\ \bibinfo {author} {\bibfnamefont
  {W.}~\bibnamefont {Hu}},\ }\bibfield  {title} {\bibinfo {title} {Chaotic
  optical communication over 1000 km transmission by coherent detection},\
  }\href@noop {} {\bibfield  {journal} {\bibinfo  {journal} {Journal of
  Lightwave Technology}\ }\textbf {\bibinfo {volume} {38}},\ \bibinfo {pages}
  {4648} (\bibinfo {year} {2020})}\BibitemShut {NoStop}%
\bibitem [{\citenamefont {Roque}\ \emph {et~al.}(2020)\citenamefont {Roque},
  \citenamefont {Marquardt},\ and\ \citenamefont
  {Yevtushenko}}]{roque2020nonlinear}%
  \BibitemOpen
  \bibfield  {author} {\bibinfo {author} {\bibfnamefont {T.~F.}\ \bibnamefont
  {Roque}}, \bibinfo {author} {\bibfnamefont {F.}~\bibnamefont {Marquardt}},\
  and\ \bibinfo {author} {\bibfnamefont {O.~M.}\ \bibnamefont {Yevtushenko}},\
  }\bibfield  {title} {\bibinfo {title} {Nonlinear dynamics of weakly
  dissipative optomechanical systems},\ }\href@noop {} {\bibfield  {journal}
  {\bibinfo  {journal} {New Journal of Physics}\ }\textbf {\bibinfo {volume}
  {22}},\ \bibinfo {pages} {013049} (\bibinfo {year} {2020})}\BibitemShut
  {NoStop}%
\bibitem [{\citenamefont {Bakemeier}\ \emph {et~al.}(2015)\citenamefont
  {Bakemeier}, \citenamefont {Alvermann},\ and\ \citenamefont
  {Fehske}}]{bakemeier2015route}%
  \BibitemOpen
  \bibfield  {author} {\bibinfo {author} {\bibfnamefont {L.}~\bibnamefont
  {Bakemeier}}, \bibinfo {author} {\bibfnamefont {A.}~\bibnamefont
  {Alvermann}},\ and\ \bibinfo {author} {\bibfnamefont {H.}~\bibnamefont
  {Fehske}},\ }\bibfield  {title} {\bibinfo {title} {Route to chaos in
  optomechanics},\ }\href@noop {} {\bibfield  {journal} {\bibinfo  {journal}
  {Physical review letters}\ }\textbf {\bibinfo {volume} {114}},\ \bibinfo
  {pages} {013601} (\bibinfo {year} {2015})}\BibitemShut {NoStop}%
\bibitem [{\citenamefont {Zhang}\ \emph {et~al.}(2012)\citenamefont {Zhang},
  \citenamefont {Wiederhecker}, \citenamefont {Manipatruni}, \citenamefont
  {Barnard}, \citenamefont {McEuen},\ and\ \citenamefont
  {Lipson}}]{zhang2012synchronization}%
  \BibitemOpen
  \bibfield  {author} {\bibinfo {author} {\bibfnamefont {M.}~\bibnamefont
  {Zhang}}, \bibinfo {author} {\bibfnamefont {G.~S.}\ \bibnamefont
  {Wiederhecker}}, \bibinfo {author} {\bibfnamefont {S.}~\bibnamefont
  {Manipatruni}}, \bibinfo {author} {\bibfnamefont {A.}~\bibnamefont
  {Barnard}}, \bibinfo {author} {\bibfnamefont {P.}~\bibnamefont {McEuen}},\
  and\ \bibinfo {author} {\bibfnamefont {M.}~\bibnamefont {Lipson}},\
  }\bibfield  {title} {\bibinfo {title} {Synchronization of micromechanical
  oscillators using light},\ }\href@noop {} {\bibfield  {journal} {\bibinfo
  {journal} {Physical review letters}\ }\textbf {\bibinfo {volume} {109}},\
  \bibinfo {pages} {233906} (\bibinfo {year} {2012})}\BibitemShut {NoStop}%
\bibitem [{\citenamefont {Koehler}\ \emph {et~al.}(2018)\citenamefont
  {Koehler}, \citenamefont {Chevalier}, \citenamefont {Shim}, \citenamefont
  {Desiatov}, \citenamefont {Shams-Ansari}, \citenamefont {Piccardo},
  \citenamefont {Okawachi}, \citenamefont {Yu}, \citenamefont {Loncar},
  \citenamefont {Lipson} \emph {et~al.}}]{koehler2018direct}%
  \BibitemOpen
  \bibfield  {author} {\bibinfo {author} {\bibfnamefont {L.}~\bibnamefont
  {Koehler}}, \bibinfo {author} {\bibfnamefont {P.}~\bibnamefont {Chevalier}},
  \bibinfo {author} {\bibfnamefont {E.}~\bibnamefont {Shim}}, \bibinfo {author}
  {\bibfnamefont {B.}~\bibnamefont {Desiatov}}, \bibinfo {author}
  {\bibfnamefont {A.}~\bibnamefont {Shams-Ansari}}, \bibinfo {author}
  {\bibfnamefont {M.}~\bibnamefont {Piccardo}}, \bibinfo {author}
  {\bibfnamefont {Y.}~\bibnamefont {Okawachi}}, \bibinfo {author}
  {\bibfnamefont {M.}~\bibnamefont {Yu}}, \bibinfo {author} {\bibfnamefont
  {M.}~\bibnamefont {Loncar}}, \bibinfo {author} {\bibfnamefont
  {M.}~\bibnamefont {Lipson}}, \emph {et~al.},\ }\bibfield  {title} {\bibinfo
  {title} {Direct thermo-optical tuning of silicon microresonators for the
  mid-infrared},\ }\href@noop {} {\bibfield  {journal} {\bibinfo  {journal}
  {Optics express}\ }\textbf {\bibinfo {volume} {26}},\ \bibinfo {pages}
  {34965} (\bibinfo {year} {2018})}\BibitemShut {NoStop}%
\bibitem [{\citenamefont {Gauthier}\ and\ \citenamefont
  {Bienfang}(1996)}]{gauthier1996intermittent}%
  \BibitemOpen
  \bibfield  {author} {\bibinfo {author} {\bibfnamefont {D.~J.}\ \bibnamefont
  {Gauthier}}\ and\ \bibinfo {author} {\bibfnamefont {J.~C.}\ \bibnamefont
  {Bienfang}},\ }\bibfield  {title} {\bibinfo {title} {Intermittent loss of
  synchronization in coupled chaotic oscillators: Toward a new criterion for
  high-quality synchronization},\ }\href@noop {} {\bibfield  {journal}
  {\bibinfo  {journal} {Physical Review Letters}\ }\textbf {\bibinfo {volume}
  {77}},\ \bibinfo {pages} {1751} (\bibinfo {year} {1996})}\BibitemShut
  {NoStop}%
\bibitem [{\citenamefont {Pesin}(1977)}]{pesin1977characteristic}%
  \BibitemOpen
  \bibfield  {author} {\bibinfo {author} {\bibfnamefont {Y.~B.}\ \bibnamefont
  {Pesin}},\ }\bibfield  {title} {\bibinfo {title} {Characteristic lyapunov
  exponents and smooth ergodic theory},\ }\href@noop {} {\bibfield  {journal}
  {\bibinfo  {journal} {Russian Mathematical Surveys}\ }\textbf {\bibinfo
  {volume} {32}},\ \bibinfo {pages} {55} (\bibinfo {year} {1977})}\BibitemShut
  {NoStop}%
\bibitem [{\citenamefont {Lu}\ \emph {et~al.}(2005)\citenamefont {Lu},
  \citenamefont {Yang}, \citenamefont {Oh},\ and\ \citenamefont
  {Luo}}]{LU20051879}%
  \BibitemOpen
  \bibfield  {author} {\bibinfo {author} {\bibfnamefont {J.}~\bibnamefont
  {Lu}}, \bibinfo {author} {\bibfnamefont {G.}~\bibnamefont {Yang}}, \bibinfo
  {author} {\bibfnamefont {H.}~\bibnamefont {Oh}},\ and\ \bibinfo {author}
  {\bibfnamefont {A.~C.}\ \bibnamefont {Luo}},\ }\bibfield  {title} {\bibinfo
  {title} {Computing lyapunov exponents of continuous dynamical systems: method
  of lyapunov vectors},\ }\href@noop {} {\bibfield  {journal} {\bibinfo
  {journal} {Chaos, Solitons \& Fractals}\ }\textbf {\bibinfo {volume} {23}},\
  \bibinfo {pages} {1879} (\bibinfo {year} {2005})}\BibitemShut {NoStop}%
\bibitem [{\citenamefont {Baptista}\ \emph {et~al.}(2012)\citenamefont
  {Baptista}, \citenamefont {Rubinger}, \citenamefont {Viana}, \citenamefont
  {Sartorelli}, \citenamefont {Parlitz},\ and\ \citenamefont
  {Grebogi}}]{baptista2012mutual}%
  \BibitemOpen
  \bibfield  {author} {\bibinfo {author} {\bibfnamefont {M.~S.}\ \bibnamefont
  {Baptista}}, \bibinfo {author} {\bibfnamefont {R.~M.}\ \bibnamefont
  {Rubinger}}, \bibinfo {author} {\bibfnamefont {E.~R.}\ \bibnamefont {Viana}},
  \bibinfo {author} {\bibfnamefont {J.~C.}\ \bibnamefont {Sartorelli}},
  \bibinfo {author} {\bibfnamefont {U.}~\bibnamefont {Parlitz}},\ and\ \bibinfo
  {author} {\bibfnamefont {C.}~\bibnamefont {Grebogi}},\ }\bibfield  {title}
  {\bibinfo {title} {Mutual information rate and bounds for it},\ }\href@noop
  {} {\bibfield  {journal} {\bibinfo  {journal} {PLoS One}\ }\textbf {\bibinfo
  {volume} {7}},\ \bibinfo {pages} {e46745} (\bibinfo {year}
  {2012})}\BibitemShut {NoStop}%
\bibitem [{\citenamefont {Gallas}(1993)}]{gallas1993structure}%
  \BibitemOpen
  \bibfield  {author} {\bibinfo {author} {\bibfnamefont {J.~A.}\ \bibnamefont
  {Gallas}},\ }\bibfield  {title} {\bibinfo {title} {Structure of the parameter
  space of the h{\'e}non map},\ }\href@noop {} {\bibfield  {journal} {\bibinfo
  {journal} {Physical review letters}\ }\textbf {\bibinfo {volume} {70}},\
  \bibinfo {pages} {2714} (\bibinfo {year} {1993})}\BibitemShut {NoStop}%
\bibitem [{\citenamefont {Maranh{\~a}o}\ \emph {et~al.}(2008)\citenamefont
  {Maranh{\~a}o}, \citenamefont {Baptista}, \citenamefont {Sartorelli},\ and\
  \citenamefont {Caldas}}]{maranhao2008experimental}%
  \BibitemOpen
  \bibfield  {author} {\bibinfo {author} {\bibfnamefont {D.}~\bibnamefont
  {Maranh{\~a}o}}, \bibinfo {author} {\bibfnamefont {M.~S.}\ \bibnamefont
  {Baptista}}, \bibinfo {author} {\bibfnamefont {J.}~\bibnamefont
  {Sartorelli}},\ and\ \bibinfo {author} {\bibfnamefont {I.~L.}\ \bibnamefont
  {Caldas}},\ }\bibfield  {title} {\bibinfo {title} {Experimental observation
  of a complex periodic window},\ }\href@noop {} {\bibfield  {journal}
  {\bibinfo  {journal} {Physical Review E—Statistical, Nonlinear, and Soft
  Matter Physics}\ }\textbf {\bibinfo {volume} {77}},\ \bibinfo {pages}
  {037202} (\bibinfo {year} {2008})}\BibitemShut {NoStop}%
\bibitem [{\citenamefont {Medeiros}\ \emph {et~al.}(2010)\citenamefont
  {Medeiros}, \citenamefont {De~Souza}, \citenamefont {Medrano-T},\ and\
  \citenamefont {Caldas}}]{medeiros2010periodic}%
  \BibitemOpen
  \bibfield  {author} {\bibinfo {author} {\bibfnamefont {E.}~\bibnamefont
  {Medeiros}}, \bibinfo {author} {\bibfnamefont {S.}~\bibnamefont {De~Souza}},
  \bibinfo {author} {\bibfnamefont {R.}~\bibnamefont {Medrano-T}},\ and\
  \bibinfo {author} {\bibfnamefont {I.~L.}\ \bibnamefont {Caldas}},\ }\bibfield
   {title} {\bibinfo {title} {Periodic window arising in the parameter space of
  an impact oscillator},\ }\href@noop {} {\bibfield  {journal} {\bibinfo
  {journal} {Physics Letters A}\ }\textbf {\bibinfo {volume} {374}},\ \bibinfo
  {pages} {2628} (\bibinfo {year} {2010})}\BibitemShut {NoStop}%
\bibitem [{\citenamefont {de~Souza}\ \emph {et~al.}(2024)\citenamefont
  {de~Souza}, \citenamefont {Batista}, \citenamefont {Medrano-T},\ and\
  \citenamefont {Caldas}}]{de2024quasiperiodic}%
  \BibitemOpen
  \bibfield  {author} {\bibinfo {author} {\bibfnamefont {S.~L.}\ \bibnamefont
  {de~Souza}}, \bibinfo {author} {\bibfnamefont {A.~M.}\ \bibnamefont
  {Batista}}, \bibinfo {author} {\bibfnamefont {R.~O.}\ \bibnamefont
  {Medrano-T}},\ and\ \bibinfo {author} {\bibfnamefont {I.~L.}\ \bibnamefont
  {Caldas}},\ }\bibfield  {title} {\bibinfo {title} {Quasiperiodic
  shrimp-shaped domains in intrinsically coupled oscillators},\ }\href@noop {}
  {\bibfield  {journal} {\bibinfo  {journal} {Chaos: An Interdisciplinary
  Journal of Nonlinear Science}\ }\textbf {\bibinfo {volume} {34}} (\bibinfo
  {year} {2024})}\BibitemShut {NoStop}%
\bibitem [{\citenamefont {Medrano-T}\ and\ \citenamefont
  {Caldas}(2010)}]{medrano2010periodic}%
  \BibitemOpen
  \bibfield  {author} {\bibinfo {author} {\bibfnamefont {R.}~\bibnamefont
  {Medrano-T}}\ and\ \bibinfo {author} {\bibfnamefont {I.}~\bibnamefont
  {Caldas}},\ }\bibfield  {title} {\bibinfo {title} {Periodic windows
  distribution resulting from homoclinic bifurcations in the two-parameter
  space},\ }\href@noop {} {\bibfield  {journal} {\bibinfo  {journal} {arXiv
  preprint arXiv:1012.2241}\ } (\bibinfo {year} {2010})}\BibitemShut {NoStop}%
\bibitem [{\citenamefont {Grebogi}\ \emph {et~al.}(1983)\citenamefont
  {Grebogi}, \citenamefont {Ott},\ and\ \citenamefont
  {Yorke}}]{grebogi1983crises}%
  \BibitemOpen
  \bibfield  {author} {\bibinfo {author} {\bibfnamefont {C.}~\bibnamefont
  {Grebogi}}, \bibinfo {author} {\bibfnamefont {E.}~\bibnamefont {Ott}},\ and\
  \bibinfo {author} {\bibfnamefont {J.~A.}\ \bibnamefont {Yorke}},\ }\bibfield
  {title} {\bibinfo {title} {Crises, sudden changes in chaotic attractors, and
  transient chaos},\ }\href@noop {} {\bibfield  {journal} {\bibinfo  {journal}
  {Physica D: Nonlinear Phenomena}\ }\textbf {\bibinfo {volume} {7}},\ \bibinfo
  {pages} {181} (\bibinfo {year} {1983})}\BibitemShut {NoStop}%
\bibitem [{\citenamefont {Baptista}\ and\ \citenamefont
  {Kurths}(2005)}]{baptista2005chaotic}%
  \BibitemOpen
  \bibfield  {author} {\bibinfo {author} {\bibfnamefont {M.}~\bibnamefont
  {Baptista}}\ and\ \bibinfo {author} {\bibfnamefont {J.}~\bibnamefont
  {Kurths}},\ }\bibfield  {title} {\bibinfo {title} {Chaotic channel},\
  }\href@noop {} {\bibfield  {journal} {\bibinfo  {journal} {Physical Review
  E—Statistical, Nonlinear, and Soft Matter Physics}\ }\textbf {\bibinfo
  {volume} {72}},\ \bibinfo {pages} {045202} (\bibinfo {year}
  {2005})}\BibitemShut {NoStop}%
\bibitem [{\citenamefont {Baptista}\ and\ \citenamefont
  {Kurths}(2008)}]{baptista2008transmission}%
  \BibitemOpen
  \bibfield  {author} {\bibinfo {author} {\bibfnamefont {M.}~\bibnamefont
  {Baptista}}\ and\ \bibinfo {author} {\bibfnamefont {J.}~\bibnamefont
  {Kurths}},\ }\bibfield  {title} {\bibinfo {title} {Transmission of
  information in active networks},\ }\href@noop {} {\bibfield  {journal}
  {\bibinfo  {journal} {Physical Review E—Statistical, Nonlinear, and Soft
  Matter Physics}\ }\textbf {\bibinfo {volume} {77}},\ \bibinfo {pages}
  {026205} (\bibinfo {year} {2008})}\BibitemShut {NoStop}%
\bibitem [{\citenamefont {Honari}\ \emph {et~al.}(2021)\citenamefont {Honari},
  \citenamefont {Haque},\ and\ \citenamefont {Lu}}]{honari2021fabrication}%
  \BibitemOpen
  \bibfield  {author} {\bibinfo {author} {\bibfnamefont {S.}~\bibnamefont
  {Honari}}, \bibinfo {author} {\bibfnamefont {S.}~\bibnamefont {Haque}},\ and\
  \bibinfo {author} {\bibfnamefont {T.}~\bibnamefont {Lu}},\ }\bibfield
  {title} {\bibinfo {title} {Fabrication of ultra-high q silica microdisk using
  chemo-mechanical polishing},\ }\href@noop {} {\bibfield  {journal} {\bibinfo
  {journal} {Applied Physics Letters}\ }\textbf {\bibinfo {volume} {119}}
  (\bibinfo {year} {2021})}\BibitemShut {NoStop}%
\bibitem [{\citenamefont {Wu}\ \emph {et~al.}(2020)\citenamefont {Wu},
  \citenamefont {Wang}, \citenamefont {Yang}, \citenamefont {Ji}, \citenamefont
  {Shen}, \citenamefont {Bao}, \citenamefont {Gao},\ and\ \citenamefont
  {Vahala}}]{wu2020greater}%
  \BibitemOpen
  \bibfield  {author} {\bibinfo {author} {\bibfnamefont {L.}~\bibnamefont
  {Wu}}, \bibinfo {author} {\bibfnamefont {H.}~\bibnamefont {Wang}}, \bibinfo
  {author} {\bibfnamefont {Q.}~\bibnamefont {Yang}}, \bibinfo {author}
  {\bibfnamefont {Q.-x.}\ \bibnamefont {Ji}}, \bibinfo {author} {\bibfnamefont
  {B.}~\bibnamefont {Shen}}, \bibinfo {author} {\bibfnamefont {C.}~\bibnamefont
  {Bao}}, \bibinfo {author} {\bibfnamefont {M.}~\bibnamefont {Gao}},\ and\
  \bibinfo {author} {\bibfnamefont {K.}~\bibnamefont {Vahala}},\ }\bibfield
  {title} {\bibinfo {title} {Greater than one billion q factor for on-chip
  microresonators},\ }\href@noop {} {\bibfield  {journal} {\bibinfo  {journal}
  {Optics Letters}\ }\textbf {\bibinfo {volume} {45}},\ \bibinfo {pages} {5129}
  (\bibinfo {year} {2020})}\BibitemShut {NoStop}%
\bibitem [{\citenamefont {Gu}\ \emph {et~al.}(2021)\citenamefont {Gu},
  \citenamefont {Liu}, \citenamefont {Bai}, \citenamefont {Wang}, \citenamefont
  {Cheng}, \citenamefont {Li}, \citenamefont {Zhang}, \citenamefont {Li},
  \citenamefont {Shi}, \citenamefont {Xiao} \emph {et~al.}}]{gu2021dry}%
  \BibitemOpen
  \bibfield  {author} {\bibinfo {author} {\bibfnamefont {J.}~\bibnamefont
  {Gu}}, \bibinfo {author} {\bibfnamefont {J.}~\bibnamefont {Liu}}, \bibinfo
  {author} {\bibfnamefont {Z.}~\bibnamefont {Bai}}, \bibinfo {author}
  {\bibfnamefont {H.}~\bibnamefont {Wang}}, \bibinfo {author} {\bibfnamefont
  {X.}~\bibnamefont {Cheng}}, \bibinfo {author} {\bibfnamefont
  {G.}~\bibnamefont {Li}}, \bibinfo {author} {\bibfnamefont {M.}~\bibnamefont
  {Zhang}}, \bibinfo {author} {\bibfnamefont {X.}~\bibnamefont {Li}}, \bibinfo
  {author} {\bibfnamefont {Q.}~\bibnamefont {Shi}}, \bibinfo {author}
  {\bibfnamefont {M.}~\bibnamefont {Xiao}}, \emph {et~al.},\ }\bibfield
  {title} {\bibinfo {title} {Dry-etched ultrahigh-q silica microdisk resonators
  on a silicon chip},\ }\href@noop {} {\bibfield  {journal} {\bibinfo
  {journal} {Photonics Research}\ }\textbf {\bibinfo {volume} {9}},\ \bibinfo
  {pages} {722} (\bibinfo {year} {2021})}\BibitemShut {NoStop}%
\end{thebibliography}
%

\end{document}